\documentclass[12pt]{article}
\usepackage[margin=1in]{geometry}
\usepackage{amsmath}
\usepackage{enumerate}
\usepackage{amsthm}
\usepackage{authblk}
\usepackage{adjustbox}
\usepackage{graphicx}
\usepackage{blkarray}
\usepackage[backend=biber,style=apa,citestyle=authoryear,natbib=true]{biblatex}
\usepackage{amssymb}
\usepackage{float}
\usepackage{commath} 
\usepackage{mathrsfs}
\usepackage{csquotes}
\usepackage{tikz-cd}
\usepackage{algorithm}
\usepackage{algpseudocode}
\usepackage{subcaption}
\usepackage{mathtools}
\usepackage[english]{babel}

\newtheorem{theorem}{Theorem}[section]

\newtheorem{prop}{Proposition}[section]
\newtheorem{cor}{Corollary}[section]

\title{Asian option valuation under price impact}
\author[1]{Priyanshu Tiwari\thanks{tiwari.priyanshu.iitk@gmail.com}}
\author[2]{Sourav Majumdar\thanks{souravm@iitk.ac.in}}
\affil[1]{Department of Materials Science And Engineering, Indian Institute of Technology Kanpur, India \& BNY, Pune}
\affil[2]{Department of Management Sciences, Indian Institute of Technology Kanpur, India}
\date{}

\begin{document}

\maketitle

\abstract{We develop a model for arithmetic and geometric Asian option
valuation with external order flow and dealer execution costs. We first formulate it as a discrete model and then obtain its continuous time scaling limit. We derive
an explicit geometric Asian option formula under a consistent price
system and formulate the dealer's bid and ask through utility
indifference with constrained stock trading. We identify sufficient
conditions for finite reservation prices and prove absence of arbitrage
under the dealer's stock trading rules. Increasing initial inventory lowers both the reservation bid and
the ask in the cases studied. In the finite-step calculations, higher
temporary costs and greater risk aversion widen the spread for both
payoffs, with a smaller change in the midpoint of the quotes.}


\section{Introduction}
\label{sec:introduction}

In physical copper contracts, the price paid for delivery is often
based on an average of daily settlement prices. Asian options on metals for instance are traded on the London Metal Exchange
\citep{LMEAveraging2017}. For a dealer who is writing an Asian option, the cost of hedging the option takes shape over a month. During the month the dealer knows the average to date and therefore has to decide how  to adjust the hedge for the remaining days in the contract. In equity markets,
\citet{BouchaudGefenPottersWyart2004} note that the directions
of successive market orders are correlated. We model the effect of
recent net buying or selling on subsequent returns. Net buying may raise the market price for only a few days. Even if the price subsequently falls back, those higher prices still count towards the monthly average. For a dealer buying shares to hedge the call, the timing of those purchases also matters. \citet{AlmgrenThumHauptmannLi2005} find that temporary impact increases with the execution rate in institutional equity trades. Buying slowly reduces this cost but can leave the dealer exposed to price changes until the shares required to hedge are acquired.

In our model, external order flow affects the prices used to calculate
the Asian payoff. The call payoff depends on market prices which evolve independently of the dealer’s trades. The dealer accounts for the impact of its trades on execution prices when determining how much to charge for writing the call. Two dealers writing the same call owe the same payoff on any given
market price path, but their hedging costs can differ. A dealer that already holds shares can use them towards
its hedge. A dealer starting with fewer shares may need to buy more
before maturity and may require a higher price to write the call.

The frictionless geometric Asian option formula is due to
\citet{KemnaVorst1990}. \citet{Curran1994} approximates
arithmetic Asian option prices by conditioning on the geometric average.
\citet{RogersShi1995} consider a scaling argument to combine the current stock
price and its running integral into a single state variable, so that pricing
an arithmetic Asian option requires solving a PDE in that variable and time.
\citet{BoylePotapchik2008} compare numerical methods for Asian
option prices and sensitivities. Price formation through order flow is studied by \citet{Kyle1985}
and \citet{BouchaudGefenPottersWyart2004}.
\citet{BertsimasLo1998} study how a trader should divide a large order across several periods when each trade affects its execution price. They arrive at the optimal strategy by minimizing the expected cost of completing the order by a fixed deadline.
\citet{AlmgrenChriss2001} incorporate the trade-off between
execution costs and price risk.
\citet{ObizhaevaWang2013} consider a model with liquidity recovery.
\citet{HubermanStanzl2004} and \citet{Gatheral2010} examine
restrictions on price impact that rule out profitable round trips.

\citet{CetinJarrowProtter2004} study the execution price
when it depends on size of the trade using a stochastic supply curve.
\citet{CetinSonerTouzi2010} study superhedging prices
under liquidity costs.
\citet{Loeper2018} derive a nonlinear pricing equation for
exact replication in the presence of linear market impact.
\citet{BechererBilarev2024} study transient multiplicative impact
and the distinction between cash and physical settlement.
In \citet{BouchardTan2022}, the hedger's trades have a permanent
effect on the underlying price path that determines the payoff.
They solve a dual control problem from which they obtain a portfolio that replicates the payoff. In our model,
external order flow affects the prices used in the Asian payoff.
The dealer's trades affect its own execution prices and leave that
average unchanged. We derive the dealer's reservation bid and ask under trading costs
and a bound on the trading rate.

Utility indifference is studied in \citet{DavisPanasZariphopoulou1993}
for option pricing with transaction costs. Following \citet{GueantPu2017}, we price the option by utility
indifference, account for execution costs on stock trades and charge
for liquidation at maturity. \citet{EkrenNadtochiy2022} also use
exponential utility. For Lipschitz payoffs with linear temporary impact, they prove regularity of the value function
and characterize the optimal hedge. In our problem, the
dealer must also account for the average to date and the effect of
earlier trades on its execution price. Bounds on inventory and trading
rate limit how quickly it can adjust the hedge. \citet{GueantPuRoyer2015} use utility indifference to value an
accelerated share repurchase contract. The bank chooses both its
purchases and the date at which the averaging period ends. Our dealer
also trades against exposure to an average price, but the contract
fixes the averaging period in advance. The dealer chooses its stock
hedge for an arithmetic or geometric call settled in cash at maturity. \citet{LiZhangLiu2018} price Asian options with a deterministic liquidity parameter and obtain explicit
formulas for geometric options. Our exogenous geometric option formula
conditions on the current order flow state, which affects expected
returns over the remaining averaging period. We then use the same payoff to study the bid and ask
of a dealer facing execution costs.

We compare two valuations of the same arithmetic or geometric Asian
payoff. The first uses a pricing measure consistent with the market
bid and ask \citep{jouini1995martingales,guasoni2008consistent} and admits an explicit formula for geometric averaging.
The second is the price at which a dealer is willing to buy or sell
the option given its stock holding and trading constraints. We can
therefore examine how external order flow affects the market valuation
and how the dealer's quotes change with inventory at a fixed market
valuation.

The main results are as follows.
\begin{enumerate}
\item We prove convergence of the discrete market processes
$(S_t^{(N)},I_t^{(N)},R_t^{(N)})_{0\leq t\leq T}$ to their continuous
time limit, including convergence of expected Asian payoffs. Under a
sufficient spread condition, we obtain an explicit consistent price
system and derive an arithmetic Asian option pricing equation and a
geometric Asian option formula. The dependence on initial order flow and on the correlation of price
and order flow shocks is explicit in the formula for this pricing measure.

\item For the dealer's hedge, we identify sufficient
conditions on initial inventory and trading capacity under which both
reservation prices are finite. We justify dynamic programming by
applying a general weak control result and derive the associated
formal Hamilton--Jacobi--Bellman equation. 

\item We prove absence of arbitrage for stock trading from zero
initial endowment under the dealer's execution and terminal liquidation
rules. Removing the dealer's impact coefficients alone leaves temporary
costs, terminal liquidation costs and trading constraints in place.
We specify the conditions under which removing these frictions and
restrictions recovers the replication price.
For the same payoff and averaging convention, this is the Kemna--Vorst
value for a geometric Asian option and the corresponding frictionless
value for an arithmetic Asian option.

\item In the numerical experiments, stronger initial buying raises
the exogenous prices of both calls by similar percentages
(Table \ref{tab:exog-geom}). Increasing the dealer's initial stock
holding from one share to two lowers both its bid and its ask
(Table \ref{tab:num-correctness}). In the finite-step calculations
from zero inventory, higher temporary execution costs or greater risk
aversion widen the spread for both payoffs, with a much smaller change
in the midpoint of the dealer's quotes
(Table \ref{tab:spread-sens}).
\end{enumerate}

Section \ref{sec:model} introduces the discrete market model and its
continuous time limit. Section \ref{sec:exogenous} develops exogenous
valuation under a consistent price system.
Section \ref{sec:endogenous} studies the dealer's reservation prices,
control equation and stock trading rules.
Section \ref{sec:numerics} presents the numerical results. Section \ref{sec:discussion} concludes the article.

\section{A Market Microstructure Model}
\label{sec:model}

We model the underlying asset over the option's life $[0,T]$, with $T>0$.
The prices are observed at
\[
 t_m=m\Delta t,\qquad \Delta t=T/N,\qquad m=0,\ldots,N.
\]
At time $t_m$, the best bid and ask quotes are $B_m$ and $A_m$,
with $0<B_m\leq A_m$. The market price is their geometric midpoint,
\[
 S_m=\sqrt{A_mB_m}.
\]
The Asian payoff is calculated from $S_m$. External trading affects $S_m$. The dealer takes this price as given
when choosing its trading strategy. Purchases and sales occur at the ask and bid quotes respectively.

The log spread of the bid and ask quotes is defined as,
$$
l_m:=\log\left(\frac{A_m}{B_m}\right)
$$

The quotes can be parametrized as $A_m=S_me^{l_m/2}$ and
$B_m=S_me^{-l_m/2}$. External net trading during $[t_m,t_{m+1})$
has direction $\varepsilon_m\in\{-1,1\}$ and volume $v_m\geq0$.
The values $+1$ and $-1$ correspond to net buying and net selling.
Past external trading affects market returns through the state $I_m$.
Trading during $[t_m,t_{m+1})$ contributes
\[
q_m=\varepsilon_m v_m^\psi
\]
to the next state:
\[
I_{m+1}=\alpha I_m+q_m,\qquad 0<\alpha<1.
\]
Each interval reduces the weight of an earlier trade by a factor
of $\alpha$. The exponent $\psi\in(0,1]$ allows a concave dependence
on volume, including the linear case $\psi=1$
\citep{bouchaud2010}.
It therefore follows,
\begin{equation}
	I_m=\alpha^m I_0+\sum_{j=0}^{m-1}\alpha^{m-1-j}q_j
\end{equation}

A larger $\alpha$ means slower decay.

$\xi_{m+1}\in\{-1,1\}$ determines the direction of the random price shock during $[t_m,t_{m+1}]$. $(\xi_{m+1},\varepsilon_m)$ at $t_{m+1}$ has the joint distribution
$$
\mathbb{P}(\xi_{m+1}=\xi,\varepsilon_m=\varepsilon)
=
\frac{1+\rho\xi\varepsilon}{4},
\qquad
\xi,\varepsilon\in\{-1,1\},
$$

where $\rho\in[-1,1]$. We assume that the pairs
$(\xi_{m+1},\varepsilon_m)$ are independent across $m$.
The information available at $t_m$ is
\[
 \mathcal F_m=\sigma\bigl((\xi_{j+1},\varepsilon_j):0\leq j<m\bigr).
\]
$\rho$ is the correlation between
$\xi_{m+1}$ and $\varepsilon_m$.

We specify the evolution of the market price by
\begin{equation}
S_{m+1} = S_m \exp\left\{ \left[ \mu(t_m)+\lambda_I I_m-\frac{1}{2}\sigma^2 \right]\Delta t + \sigma\sqrt{\Delta t} \xi_{m+1} \right\}, \label{eq:midpoint-discrete}
\end{equation}
where $\mu(t_m)$ is the baseline rate of return of the market price,
$\sigma>0$ is its volatility and $\lambda_I$ measures the effect of
$I_m$ on the conditional return of the market price.
A trade's contribution to subsequent log returns decays with $I_m$.
These contributions accumulate in the log price.

The Asian payoff depends on
\begin{equation}
    R_{m+1}=R_m+h(S_m)\Delta t,\qquad R_0=0,
    \label{eq:R-discrete}
\end{equation}
where $h(s)=s$ for arithmetic averaging and $h(s)=\log s$
for geometric averaging. The corresponding discrete averages are
$R_N/T$ and $\exp(R_N/T)$, with call payoffs
\[
    \Phi_A(R_N)=\left(\frac{R_N}{T}-K\right)^+,
    \qquad
    \Phi_G(R_N)=\left(e^{R_N/T}-K\right)^+.
\]

\subsection{Continuous time limit}
\label{subsec:diffusion-limit}

For each time step $\Delta t$, we choose the decay factor and trade
volume according to
\begin{equation}
    \alpha=1-\kappa_I\Delta t,\qquad
    v_m^\psi=\eta(t_m)\sqrt{\Delta t},
    \label{eq:impact-scaling}
\end{equation}
where $\kappa_I>0$ is the mean reversion rate and $\eta(t)\geq0$
controls the magnitude of random fluctuations in $I_m$.
For $N>\kappa_I T$, the condition $0<\alpha<1$ holds.
Substituting this scaling into the equation for $I_m$, we obtain
\begin{equation}
    I_{m+1}=I_m-\kappa_I I_m\Delta t
        +\eta(t_m)\sqrt{\Delta t}\varepsilon_m.
    \label{eq:impact-state-discrete}
\end{equation}

\begin{theorem}\label{thm:diffusion-limit}
For the model and scaling above, suppose $\mu$ and $\eta$ are
deterministic and continuous on $[0,T]$, $g(x)=h(e^x)$ is continuous,
and $S_0>0$.
As $N\to\infty$, the piecewise constant interpolations
$(S_t^{(N)},I_t^{(N)},R_t^{(N)})$, equal to $(S_m,I_m,R_m)$
on $[t_m,t_{m+1})$ and to $(S_N,I_N,R_N)$ at $T$, satisfy
\begin{equation}
 (S_t^{(N)},I_t^{(N)},R_t^{(N)})_{0\leq t\leq T}\Rightarrow(S_t,I_t,R_t)_{0\leq t\leq T}
 \label{eq:joint-convergence}
\end{equation}
in $D([0,T];\mathbb R^3)$ with the Skorokhod $J_1$ topology, where
$(S_t,I_t,R_t)_{0\leq t\leq T}$ is the unique strong solution,
with the same initial values, of
\begin{align}
 \frac{dS_t}{S_t}&=(\mu(t)+\lambda_I I_t) dt+\sigma dW_t,
 \label{eq:S-limit}\\
 dI_t&=-\kappa_I I_t dt+\eta(t) dW_t^I,
 \label{eq:I-limit}\\
 dR_t&=h(S_t) dt,
 \label{eq:R-limit}
\end{align}
and $W_t,W_t^I$ are Brownian motions satisfying
\begin{equation}
 d\langle W,W^I\rangle_t=\rho dt.
 \label{eq:Brownian-correlation}
\end{equation}
\end{theorem}

\begin{proof}
The driving processes are
\begin{align}
    W_t^{(N)}
    &=
    \sum_{j=0}^{\lfloor t/\Delta t\rfloor-1}
    \xi_{j+1}\sqrt{\Delta t},
    \label{eq:WN}\\
    M_t^{(N)}
    &=
    \sum_{j=0}^{\lfloor t/\Delta t\rfloor-1}
    \eta(t_j)\varepsilon_j\sqrt{\Delta t}.
    \label{eq:MN}
\end{align}

The pair $Z_t^{(N)}=(W_t^{(N)},M_t^{(N)})$ has increments
\[
    \Delta Z_j^{(N)}
    =
    \begin{pmatrix}
        \xi_{j+1}\sqrt{\Delta t}\\
        \eta(t_j)\varepsilon_j\sqrt{\Delta t}
    \end{pmatrix}.
\]
For the independent shock pairs specified above, the conditional moments are
\[
 \mathbb E[\Delta Z_j^{(N)}\mid\mathcal F_j]=0,
 \qquad
 \mathbb E[\|\Delta Z_j^{(N)}\|^2\mid\mathcal F_j]
 =\Delta t\bigl(1+\eta(t_j)^2\bigr)<\infty.
\]
Thus $(Z_t^{(N)})_{0\leq t\leq T}$ is an $L^2$ martingale for the filtration
\[
 \mathcal F_t^{(N)}=\mathcal F_{\lfloor t/\Delta t\rfloor}.
\]
Its conditional covariance matrix is
\[
    \mathbb{E}
    \left[
        \Delta Z_j^{(N)}
        \big(\Delta Z_j^{(N)}\big)^\top
        \mid\mathcal{F}_j
    \right]
    =
    \Delta t
    \begin{pmatrix}
        1 & \rho\eta(t_j)\\
        \rho\eta(t_j) & \eta(t_j)^2
    \end{pmatrix}.
\]

The predictable quadratic covariations satisfy, uniformly on $[0,T]$,
\begin{align*}
 \langle W^{(N)}\rangle_t
 &=\lfloor t/\Delta t\rfloor\Delta t\longrightarrow t,\\
 \langle M^{(N)}\rangle_t
 &=\sum_{j=0}^{\lfloor t/\Delta t\rfloor-1}\eta(t_j)^2\Delta t
 \longrightarrow\int_0^t\eta(u)^2du,\\
 \langle W^{(N)},M^{(N)}\rangle_t
 &=\rho\sum_{j=0}^{\lfloor t/\Delta t\rfloor-1}\eta(t_j)\Delta t
 \longrightarrow\rho\int_0^t\eta(u)du.
\end{align*}
Moreover,
\[
 \max_{j<N}\|\Delta Z_j^{(N)}\|^2
 \leq\Delta t(1+\|\eta\|_\infty^2)\longrightarrow0.
\]
For every $\delta>0$ and all sufficiently large $N$,
\[
 \sum_{j=0}^{N-1}\mathbb E\left[
 \|\Delta Z_j^{(N)}\|^2
 \mathbf 1_{\{\|\Delta Z_j^{(N)}\|>\delta\}}
 \mid\mathcal F_j\right]=0,
\]
which is the conditional Lindeberg condition
\citep[Theorem 8.2.4, condition (ii)]{durrett2019probability}.
The jumps of the predictable quadratic covariations obey the same
bound. By the martingale functional central limit theorem
\citep[Chapter 7, Theorem 1.4(b)]{ethier1986markov},
\begin{equation}
 (W_t^{(N)},M_t^{(N)})_{0\leq t\leq T}\Rightarrow(W_t,M_t)_{0\leq t\leq T},
 \label{eq:driving-convergence}
\end{equation}
where
\begin{equation}
 M_t=\int_0^t\eta(u) dW_u^I,
 \label{eq:M-representation}
\end{equation}
with $W_t,W_t^I$ Brownian motions satisfying
$d\langle W,W^I\rangle_t=\rho dt$.

We next compare the discrete and limiting equations for each sample
path. The Skorokhod representation theorem
\citep[Theorem 1.8]{ethier1986markov} guarantees the
existence of copies of $(W_t^{(N)},M_t^{(N)})_{0\leq t\leq T}$ and
$(W_t,M_t)_{0\leq t\leq T}$ on a common probability space,
with their respective joint distributions unchanged, such that
the former converge a.s. to the latter in $J_1$.
Since the limiting paths are continuous, the
convergence is uniform on $[0,T]$
\citep[Lemma 10.1]{ethier1986markov}:
\begin{equation}
 \|W^{(N)}-W\|_\infty+\|M^{(N)}-M\|_\infty\longrightarrow0
 \quad\text{a.s.}
 \label{eq:M-uniform}
\end{equation}
The state equations then determine copies of
$(S_t^{(N)},I_t^{(N)},R_t^{(N)})_{0\leq t\leq T}$
with the original joint distribution.

The most recent observation time is
$\tau_N(t)=\Delta t\lfloor t/\Delta t\rfloor$.
Equation \eqref{eq:impact-state-discrete} and its candidate limit are
\begin{align}
 I_t^{(N)}&=I_0-\kappa_I\int_0^{\tau_N(t)}I_u^{(N)} du+M_t^{(N)},
 \label{eq:I-integral-discrete}\\
 I_t&=I_0-\kappa_I\int_0^t I_u du+M_t.
 \label{eq:I-integral-limit}
\end{align}
Since $0\leq t-\tau_N(t)\leq\Delta t$, from Gronwall's
inequality we obtain
\begin{equation}
 \|I^{(N)}-I\|_\infty
 \leq e^{\kappa_I T}\left(
 \|M^{(N)}-M\|_\infty+\kappa_I\Delta t \|I\|_\infty
 \right)\longrightarrow0
 \quad\text{a.s.}
 \label{eq:I-uniform}
\end{equation}

The log price $X_t^{(N)}=\log S_t^{(N)}$ satisfies
\begin{equation}
 X_t^{(N)}=X_0+
 \int_0^{\tau_N(t)}
 \left[\mu(\tau_N(u))+\lambda_I I_u^{(N)}-\frac12\sigma^2\right]du
 +\sigma W_t^{(N)},
 \label{eq:X-integral-discrete}
\end{equation}
where $X_0=\log S_0$. Its candidate limit is
\begin{equation}
 X_t=X_0+\int_0^t
 \left[\mu(u)+\lambda_I I_u-\frac12\sigma^2\right]du+\sigma W_t.
 \label{eq:X-integral-limit}
\end{equation}
By continuity of $\mu$, \eqref{eq:M-uniform} and
\eqref{eq:I-uniform},
\begin{equation}
 \begin{aligned}
 \|X^{(N)}-X\|_\infty
 \leq{}&T\|\mu\circ\tau_N-\mu\|_\infty
       +|\lambda_I|T\|I^{(N)}-I\|_\infty\\
 &+\sigma\|W^{(N)}-W\|_\infty\\
 &+\Delta t\left(\|\mu\|_\infty
       +|\lambda_I|\|I\|_\infty+\tfrac12\sigma^2\right)
 \longrightarrow0\quad\text{a.s.}
 \end{aligned}
 \label{eq:X-uniform}
\end{equation}
For sufficiently large $N$,
$\|X^{(N)}\|_\infty\leq\|X\|_\infty+1$ a.s. Thus, for $S_t=e^{X_t}$,
\[
 \|S^{(N)}-S\|_\infty
 \leq e^{\|X\|_\infty+1}\|X^{(N)}-X\|_\infty
 \longrightarrow0\quad\text{a.s.}
\]
From It\^{o}'s formula, $S_t$ satisfies \eqref{eq:S-limit}.

For $g(x)=h(e^x)$, the discrete equation and its limit are
\begin{align}
 R_t^{(N)}&=R_0+\int_0^{\tau_N(t)}g(X_u^{(N)})du,
 \label{eq:R-integral-discrete}\\
 R_t&=R_0+\int_0^t g(X_u)du.
 \label{eq:R-integral-limit}
\end{align}
By uniform continuity of $g$ on
$[-\|X\|_\infty-1,\|X\|_\infty+1]$,
\begin{equation}
 \|R^{(N)}-R\|_\infty
 \leq T\sup_{0\leq u\leq T}|g(X_u^{(N)})-g(X_u)|
 +\Delta t\sup_{0\leq u\leq T}|g(X_u)|
 \longrightarrow0
 \quad\text{a.s.}
\end{equation}
$I_t$ has a unique strong solution, and
\eqref{eq:X-integral-limit} and \eqref{eq:R-integral-limit} uniquely
determine $S_t$ and $R_t$.
On the common probability space,
\[
 \|S^{(N)}-S\|_\infty+\|I^{(N)}-I\|_\infty
 +\|R^{(N)}-R\|_\infty\longrightarrow0\quad\text{a.s.}
\]
Uniform convergence implies $J_1$ convergence. Preservation of the
joint laws then implies \eqref{eq:joint-convergence}.

\end{proof}

\begin{cor}[Convergence of Asian payoff expectations]
\label{cor:payoff-convergence}
Under the assumptions of Theorem \ref{thm:diffusion-limit}, with
$R_0=0$ and $K>0$, either Asian call option payoff defined above satisfies
\begin{equation}
 \mathbb E[\Phi(R_N)]\longrightarrow\mathbb E[\Phi(R_T)].
 \label{eq:payoff-convergence}
\end{equation}
The same conclusion holds with a fixed deterministic discount factor.
\end{cor}
\begin{proof}
By Theorem \ref{thm:diffusion-limit} and continuity of $\Phi$,
\[
 H_N=\Phi(R_N)\Rightarrow H=\Phi(R_T).
\]
For either $U_t^{(N)}=W_t^{(N)}$ or $U_t^{(N)}=M_t^{(N)}$, independence
of the symmetric increments and $\cosh z\leq e^{z^2/2}$ imply
\[
 \mathbb E[e^{\theta U_T^{(N)}}]
 =\prod_{j=0}^{N-1}\mathbb E[e^{\theta\Delta U_j^{(N)}}]
 \leq e^{\theta^2V_0/2},
 \qquad V_0=T(1+\|\eta\|_\infty^2),\quad\theta\in\mathbb R.
\]
By Doob's maximal inequality
\citep[Theorem 4.4.2]{durrett2019probability},
\[
 \mathbb P(\|U^{(N)}\|_\infty>x)
 \leq\inf_{\theta>0}2e^{-\theta x+\theta^2V_0/2}
 =2e^{-x^2/(2V_0)},\qquad x>0.
\]
Consequently,
\begin{equation}
 \sup_N\mathbb E[e^{a\|U^{(N)}\|_\infty}]
 \leq1+2a\int_0^\infty e^{ax-x^2/(2V_0)}dx<\infty,
 \qquad a>0.
 \label{eq:noise-exponential-moments}
\end{equation}
From Gronwall's inequality, \eqref{eq:I-integral-discrete} and
\eqref{eq:X-integral-discrete},
\[
 \|X^{(N)}\|_\infty
 \leq C_0+C_1\|M^{(N)}\|_\infty+\sigma\|W^{(N)}\|_\infty,
\]
with $C_0,C_1\geq0$ independent of $N$. By Cauchy--Schwarz,
\begin{align*}
 \mathbb E[e^{p\|X^{(N)}\|_\infty}]
 &\leq e^{pC_0}
 \left(\mathbb E[e^{2pC_1\|M^{(N)}\|_\infty}]
       \mathbb E[e^{2p\sigma\|W^{(N)}\|_\infty}]\right)^{1/2}.
\end{align*}
Thus, by \eqref{eq:noise-exponential-moments},
\begin{equation}
 \sup_N\mathbb E[e^{p\|X^{(N)}\|_\infty}]<\infty,
 \qquad p>0.
 \label{eq:log-price-exponential-moments}
\end{equation}
For either Asian payoff,
\[
 0\leq H_N\leq\max_{m<N}S_m\leq e^{\|X^{(N)}\|_\infty},
 \qquad
 \sup_N\mathbb E[H_N^2]
 \leq\sup_N\mathbb E[e^{2\|X^{(N)}\|_\infty}]<\infty.
\]
Hence $(H_N)_{N\geq1}$ is uniformly integrable. From
\citet[Theorems 3.2.8 and 4.6.3]{durrett2019probability},
\[
 \mathbb E[H_N]\longrightarrow\mathbb E[H].
\]
\end{proof}

\section{Exogenous valuation}
\label{sec:exogenous}

We first price the arithmetic and geometric Asian payoffs in the
market of Section \ref{sec:model}, where external trading affects
the midpoint of bid and ask price of $S_t$. Stock purchases occur at the ask and sales at
the bid price. The valuation is \emph{exogenous} because the market
prices and the resulting option value are unaffected by the
dealer's strategy. It serves as a benchmark comparison for the reservation
prices in endogenous valuation in Section \ref{sec:endogenous}.

The first issue is to specify a pricing measure compatible with
these stock quotes. We do so through a consistent price system
and prove that the discounted expected Asian payoff can be
introduced as an option price without creating arbitrage under
the stated trading rules. We then derive the valuation equation
for the arithmetic Asian option and an explicit formula for the
geometric Asian option.

\subsection{A consistent price system}
\label{subsec:benchmark-valuation}

The Asian payoff is calculated from the midpoint of the bid and ask quotes, while stock
trades take place at the bid and ask quotes. The pricing measure must
therefore be consistent with these trading prices. In a
frictionless market, an equivalent risk-neutral measure makes
the discounted stock price a martingale. With bid and ask quotes, the
corresponding condition can be imposed on an auxiliary stock
price lying between them.

\citet[Theorem 3.2]{jouini1995martingales} showed the absence
of free lunches with transaction costs to the existence of such
martingales and used the associated measures to characterize
hedging costs.
\citet{guasoni2008consistent} studied the existence of
consistent price systems for stock prices and their
use in superreplication. The auxiliary price offers purchases
at or below the ask and sales at or above the bid. Comparing
trading gains at these prices is the basis of the no-arbitrage
argument \citep[Remark 2.2]{guasoni2008consistent}.

We seek to construct a pricing measure under which the drift of $S_t$ depends
on $I_t$. The consistent price system developed below has this
property whenever the spread satisfies the stated lower bound.

The bank account is $D_t=e^{rt}$, with constant interest rate $r$.
The stock pays no dividends. As in Section \ref{sec:model}, its bid
and ask prices are
\begin{equation}
    B_t=S_t e^{-l_t/2},\qquad A_t=S_t e^{l_t/2},
    \label{eq:continuous-quotes}
\end{equation}
where $l_t$ is a continuous adapted log spread, specified below.

A \emph{consistent price system} consists of a measure $\mathbb Q$
equivalent to $\mathbb P$ on $\mathcal F_T$\footnote{Throughout this
section, $\mathcal F_t$ is the usual augmented filtration generated
by the Brownian motions in Theorem \ref{thm:diffusion-limit}.}
and an adapted auxiliary price $\widetilde S_t$ such that
\begin{equation}
    B_t\leq\widetilde S_t\leq A_t,
    \qquad (e^{-rt}\widetilde S_t)_{0\leq t\leq T}
    \text{ is a martingale under }\mathbb Q.
    \label{eq:cps-definition}
\end{equation}
The auxiliary price is used to verify absence of arbitrage.
The Asian payoff continues to depend on the midpoint $S_t$.

Our choice of $\mathbb Q$ replaces the baseline drift $\mu(t)$
by a deterministic continuous function $\mu_{\mathbb Q}(t)$ and
leaves the dynamics of $I_t$ unchanged.  For $|\rho|<1$, the Brownian motions can be expressed as
$W_t=B_t^1$ and $W_t^I=\rho B_t^1+\sqrt{1-\rho^2}B_t^2$,
where $B_t^1,B_t^2$ are independent under $\mathbb P$.
The change of measure uses
\[
    \phi_1(t)=\frac{\mu(t)-\mu_{\mathbb Q}(t)}{\sigma},
    \qquad
    \phi_2(t)=-\frac{\rho\phi_1(t)}{\sqrt{1-\rho^2}}.
\]
The pricing measure $\mathbb Q$ is defined by
\begin{equation}
    \frac{d\mathbb Q}{d\mathbb P}\bigg|_{\mathcal F_T}
    =\exp\left\{
      -\sum_{j=1}^2\int_0^T\phi_j(u) dB_u^j
      -\frac12\int_0^T\bigl(\phi_1(u)^2+\phi_2(u)^2\bigr)du
    \right\}.
    \label{eq:pricing-density}
\end{equation}
Novikov's condition
holds and $\mathbb Q\sim\mathbb P$. From Girsanov's theorem the dynamics under $\mathbb Q$ are
\begin{align}
    \frac{dS_t}{S_t}
    &=\bigl(\mu_{\mathbb Q}(t)+\lambda_I I_t\bigr)dt
      +\sigma dW_t^{\mathbb Q},
    \label{eq:S-Q}\\
    dI_t&=-\kappa_I I_t dt+\eta(t) dW_t^{I,\mathbb Q},
    \label{eq:I-Q}\\
    dR_t&=h(S_t)dt,
    \label{eq:R-Q}
\end{align}
where
\begin{equation}
    d\langle W^{\mathbb Q},W^{I,\mathbb Q}\rangle_t=\rho dt.
    \label{eq:corr-Q}
\end{equation}
Indeed, $B_t^{j,\mathbb Q}=B_t^j+\int_0^t\phi_j(u)du$ are
independent Brownian motions, and
$\rho\phi_1(t)+\sqrt{1-\rho^2}\phi_2(t)=0$ preserves the drift
of $I_t$. For $|\rho|=1$, we require
$\eta(t)(\mu(t)-\mu_{\mathbb Q}(t))=0$ on $[0,T]$ and use the
one-dimensional density with $\phi_1$ and $W_t$ instead. 

We seek an auxiliary price of the form
$\widetilde S_t=S_t\exp\{\beta I_t+b(t)\}$ with return drift $r$.
We therefore make the following choices,
\begin{equation}
    \beta=\frac{\lambda_I}{\kappa_I},\qquad
    b(t)=\int_t^T\left[
       \mu_{\mathbb Q}(u)-r+\rho\sigma\beta\eta(u)
       +\frac12\beta^2\eta(u)^2\right]du.
    \label{eq:cps-correction}
\end{equation}
To place this price strictly inside the quotes, we require, for
some $\epsilon>0$,
\begin{equation}
    \frac{l_t}{2}\geq|\beta I_t+b(t)|+\epsilon,
    \qquad 0\leq t\leq T,\quad\text{a.s.}
    \label{eq:cps-spread}
\end{equation}
One explicit choice is $l_t=2|\beta I_t+b(t)|+2\epsilon$.
This is a sufficient spread condition for our construction.
For $\lambda_I\ne0$, unbounded possible values of $I_t$ require
unbounded possible spreads under this condition.

\begin{theorem}[A consistent price system]
\label{thm:consistent-price-system}
Under \eqref{eq:cps-spread}, the measure \eqref{eq:pricing-density}
(with the stated modification when $|\rho|=1$) and the auxiliary price
\begin{equation}
    \widetilde S_t=S_t\exp\{\beta I_t+b(t)\}
    \label{eq:auxiliary-price}
\end{equation}
form a consistent price system. In fact,
$B_t<\widetilde S_t<A_t$ for every $t\in[0,T]$, a.s.
\end{theorem}

\begin{proof}
The spread condition places \eqref{eq:auxiliary-price} strictly between
the quotes. Applying It\^{o}'s formula and using $\beta\kappa_I=\lambda_I$
and \eqref{eq:cps-correction}, we obtain
\begin{equation}
    \frac{d\widetilde S_t}{\widetilde S_t}
    =r dt+\sigma dW_t^{\mathbb Q}
          +\beta\eta(t)dW_t^{I,\mathbb Q}.
    \label{eq:auxiliary-dynamics}
\end{equation}
The term involving $I_t$ cancels from the drift. The quadratic
variation rate of the diffusion term is the deterministic bounded function
\[
    v(t)=\sigma^2+2\rho\sigma\beta\eta(t)+\beta^2\eta(t)^2\geq0.
\]
Consequently,
\[
    e^{-rt}\widetilde S_t
    =\widetilde S_0\exp\left\{
       \sigma W_t^{\mathbb Q}
       +\beta\int_0^t\eta(u)dW_u^{I,\mathbb Q}
       -\frac12\int_0^t v(u)du\right\}
\]
is a martingale. Since $\mathbb Q\sim\mathbb P$, the pair satisfies
the definition of a consistent price system in \eqref{eq:cps-definition}.
\end{proof}

For comparison with standard Asian option prices, we take
\begin{equation}
    \mu_{\mathbb Q}(t)=r.
    \label{eq:benchmark-r}
\end{equation}
The midpoint return drift is then $r+\lambda_I I_t$ and reduces
to $r$ when $\lambda_I=0$. At $\lambda_I=0$, the price can therefore be compared with the
frictionless Asian option value.
\subsection{Arbitrage free Asian option valuation}
\label{subsec:exogenous-price}

The measure $\mathbb Q$ from Section \ref{subsec:benchmark-valuation}
suggests the discounted expected payoff as an Asian option price.
For either payoff this expectation is finite. We show below that
stock trading at the quoted bid and ask remains free of arbitrage
when purchases and sales of the option at this price are also allowed.

\citet{AlmgrenLi2016} study how
market impact changes option hedging, and
\citet{BankSonerVoss2017} formulate hedging as a problem of
tracking a frictionless strategy subject to trading costs.
\citet{GueantPu2017} determine option prices through utility
indifference. We use the exogenous price as the reference for the reservation
prices calculated in Section \ref{sec:endogenous}.

The option pays $\Phi(R_T)$ in cash at maturity, with strike $K>0$
and the arithmetic or geometric payoff from Section \ref{sec:model}.
For either contract, the proposed price is
\begin{equation}
    V(t,s,i,a)
    =\mathbb E^{\mathbb Q}_{t,s,i,a}
      \left[e^{-r(T-t)}\Phi(R_T)\right],
    \qquad V_t=V(t,S_t,I_t,R_t).
    \label{eq:generic-exogenous-value}
\end{equation}
The expectation is conditional on $S_t=s$, $I_t=i$ and $R_t=a$,
with future dynamics given by \eqref{eq:S-Q}--\eqref{eq:R-Q}.
The payoff depends on past prices through $R_t$, with $h(s)=s$
for arithmetic averaging and $h(s)=\log s$ for geometric averaging.

Stock and option holdings in the following result are predictable
c\`adl\`ag processes of finite variation. Stock trades with the prices quotes
$A_t$ for purchase and $B_t$ for a sale, while both option purchases
and sales use $V_t$. The
liquidation value is the cash obtained after both positions have
been closed at the prevailing prices. Admissibility requires
a deterministic lower bound on its discounted value over $[0,T]$.
An arbitrage is an admissible strategy with zero initial capital whose
terminal liquidation value is nonnegative a.s. and positive with
positive probability.

\begin{prop}[Arbitrage-free extension]
\label{prop:asian-no-arbitrage}
Under the assumptions of Theorem \ref{thm:consistent-price-system},
\eqref{eq:generic-exogenous-value} is finite for either Asian call option.
For the admissible strategies specified above, the stock market is
free of arbitrage and remains so when the option is introduced
with buying and selling price $V_t$.
\end{prop}

\begin{proof}
The moment estimates in the proof of Corollary \ref{cor:payoff-convergence},
with $\mu_{\mathbb Q}$ in place of $\mu$, imply finite moments of every
positive order for $S^*_{t,T}=\sup_{t\leq u\leq T}S_u$.
Starting from $R_t=a$, the payoffs satisfy
\[
    0\leq\Phi_A(R_T)\leq\frac{|a|}{T}+S^*_{t,T},\qquad
    0\leq\Phi_G(R_T)\leq e^{a/T}\max\{1,S^*_{t,T}\}.
\]
Both are integrable. Moreover,
\begin{equation}
    e^{-rt}V_t
    =\mathbb E^{\mathbb Q}
       [e^{-rT}\Phi(R_T)\mid\mathcal F_t]
    \label{eq:discounted-option-martingale}
\end{equation}
is a uniformly integrable martingale, with a continuous version in the
Brownian filtration.

To rule out arbitrage, we compare wealth at the quoted prices
with wealth valued at $\widetilde S_t$. Throughout this proof,
a bar denotes multiplication by $e^{-rt}$, for example
$\bar A_t=e^{-rt}A_t=A_t/D_t$.
The stock and option holdings are $\varphi_t$ and $\chi_t$,
both zero just before time $0$. The cash balance at time $t$ is
$D_tc_t$, so $c_t$ denotes its discounted value. The cumulative
numbers of shares bought and sold by time $t$ are
$\varphi_t^{\uparrow}$ and $\varphi_t^{\downarrow}$, respectively.
Both are nondecreasing, and the net stock holding is
$\varphi_t=\varphi_t^{\uparrow}-\varphi_t^{\downarrow}$.
The cash account pays for purchases and receives sale proceeds,
so a self-financing strategy satisfies
\[
    dc_t=-\bar A_t d\varphi_t^{\uparrow}
           +\bar B_t d\varphi_t^{\downarrow}
           -\bar V_t d\chi_t.
\]
The discounted wealth valued at the auxiliary price is
$X_t=c_t+\varphi_t\bar{\widetilde S}_t+\chi_t\bar V_t$.
For zero initial capital, we obtain the following decomposition by
integration by parts:
\begin{align*}
    X_t&=M_t-K_t,\\
    M_t&=\int_0^t\varphi_{u-}d\bar{\widetilde S}_u
          +\int_0^t\chi_{u-}d\bar V_u,\\
    K_t&=\int_{[0,t]}(\bar A_u-\bar{\widetilde S}_u)
                         d\varphi_u^{\uparrow}
          +\int_{[0,t]}(\bar{\widetilde S}_u-\bar B_u)
                         d\varphi_u^{\downarrow}.
\end{align*}
The interval $[0,t]$ includes costs of initial trades. The process $M_t$
is a local $\mathbb Q$-martingale starting from zero, and $K_t$ is
nonnegative and increasing. Discounted liquidation wealth is
\[
    L_t=c_t+\varphi_t^+\bar B_t-\varphi_t^-\bar A_t+\chi_t\bar V_t
    \leq X_t.
\]
By admissibility, $L_t\geq-m$ for some deterministic $m$, hence
$M_t=X_t+K_t\geq-m$. This lower bound makes $M_t$ a supermartingale.
If $L_T\geq0$, then $0\leq L_T\leq M_T$ and
$\mathbb E^{\mathbb Q}[M_T]\leq0$, so $L_T=0$ a.s.
Equivalence of $\mathbb Q$ and $\mathbb P$ rules out arbitrage under the
physical measure as well. Setting $\chi_t=0$ proves the assertion for
the stock market alone.
\end{proof}

The comparison of trading gains follows the consistent price system
argument in \citet[Remark 2.2]{guasoni2008consistent}.
Thus $V_t$ is an arbitrage-free price for the option under these
trading rules. Additional conditions would be needed for replication
or uniqueness. Section \ref{sec:endogenous} uses different trading
opportunities: the dealer holds a given option position and chooses
a stock hedge through its execution rule, with expected utility as
the valuation criterion.

To compute $V$, we use the Markov dynamics of $(S_t,I_t,R_t)$.
For smooth $V$, we apply It\^{o}'s formula to
\eqref{eq:discounted-option-martingale} to obtain the pricing PDE.
The regularity and integrability assumptions for identifying a
solution with $V$ are specified in the proposition below. For a smooth function
$f=f(t,s,i,a)$, the spatial generator is
\begin{align}
    \mathcal L^{\mathbb Q}f
    ={}&s\bigl(\mu_{\mathbb Q}(t)+\lambda_I i\bigr)f_s
        -\kappa_I i f_i+h(s)f_a
    \nonumber\\
    &+\frac12\sigma^2s^2f_{ss}
      +\frac12\eta(t)^2f_{ii}
      +\rho\sigma s\eta(t)f_{si}.
    \label{eq:generic-generator}
\end{align}
For the selected $\mathbb Q$, the coefficients follow directly
from \eqref{eq:S-Q}--\eqref{eq:R-Q}. The quoted spread must satisfy
the consistency condition \eqref{eq:cps-spread}.

\begin{prop}[Valuation equation]
\label{prop:generic-exogenous-PDE}
Suppose $V$ is continuous up to $t=T$ and, for $t<T$, has continuous
derivatives $\partial_tV,V_a,V_s,V_i,V_{ss},V_{si},V_{ii}$. Then
\begin{equation}
    \partial_tV+\mathcal L^{\mathbb Q}V-rV=0,\qquad
    V(T,s,i,a)=\Phi(a).
    \label{eq:generic-exogenous-PDE}
\end{equation}
Conversely, a solution $F$ with this regularity and terminal condition
equals $V$ if, for every starting state, the family
\[
    \left\{e^{-r(\tau-t)}F(\tau,S_\tau,I_\tau,R_\tau):
      \tau\text{ is a stopping time with }t\leq\tau\leq T\right\}
\]
is uniformly integrable under $\mathbb Q$.
\end{prop}

\begin{proof}
We use the argument in
\citet[Propositions 5.8--5.9]
{bjork2009arbitrage}
for $(S_t,I_t,R_t)$ with generator $\mathcal L^{\mathbb Q}$.
By It\^{o}'s formula applied to the martingale in
\eqref{eq:discounted-option-martingale}, $V$ satisfies the PDE,
with terminal value \eqref{eq:generic-exogenous-value}.
For a solution $F$ of the PDE,
$e^{-r(u-t)}F(u,S_u,I_u,R_u)$ is a local martingale by the same
calculation and a martingale by the assumed uniform integrability.
Its conditional expectation at $T$, evaluated using the terminal
condition, is $V(t,s,i,a)$. Thus $F=V$.
\end{proof}

\subsection{Arithmetic Asian option}
\label{subsec:arithmetic-exogenous}

For arithmetic averaging,
\[
    Y_t=R_t=\int_0^tS_u du,\qquad
    \Phi_A(Y_T)=\left(\frac{Y_T}{T}-K\right)^+.
\]
The price supported by the selected consistent price system is
\begin{equation}
    V_A(t,s,i,y)
    =\mathbb E^{\mathbb Q}_{t,s,i,y}
       \left[e^{-r(T-t)}\left(\frac{Y_T}{T}-K\right)^+\right].
    \label{eq:VA-definition}
\end{equation}
Under the regularity conditions of
Proposition \ref{prop:generic-exogenous-PDE}, it satisfies
\begin{align}
    0={}&\partial_tV_A+s\bigl(\mu_{\mathbb Q}(t)+\lambda_I i\bigr)(V_A)_s
       -\kappa_I i(V_A)_i+s(V_A)_y
    \nonumber\\
    &+\frac12\sigma^2s^2(V_A)_{ss}
       +\frac12\eta(t)^2(V_A)_{ii}
       +\rho\sigma s\eta(t)(V_A)_{si}-rV_A,
    \label{eq:arithmetic-PDE}
\end{align}
with
\begin{equation}
    V_A(T,s,i,y)=\left(\frac{y}{T}-K\right)^+.
    \label{eq:arithmetic-terminal}
\end{equation}
A numerical solution of \eqref{eq:arithmetic-PDE}, or simulation
under $\mathbb Q$ using \eqref{eq:S-Q}--\eqref{eq:I-Q}, determines
$V_A$. The arithmetic average is generally non-lognormal. Under
\eqref{eq:cps-spread}, the price $V_A$ satisfies the no-arbitrage
conclusion of Proposition \ref{prop:asian-no-arbitrage}.

\subsection{Geometric Asian option}
\label{subsec:geometric-exogenous}

For geometric averaging,
\begin{equation}
    Z_t=R_t=\int_0^t\log S_u du,\qquad dZ_t=\log S_t dt.
    \label{eq:Z-definition}
\end{equation}
The average is $G_T=e^{Z_T/T}$ and the cash payoff is
$\Phi_G(Z_T)=(G_T-K)^+$. Its price is
\begin{equation}
    V_G(t,s,i,z)
    =\mathbb E^{\mathbb Q}_{t,s,i,z}
       \left[e^{-r(T-t)}(e^{Z_T/T}-K)^+\right].
    \label{eq:VG-definition}
\end{equation}
Under the same regularity conditions, the valuation equation is
\begin{align}
    0={}&\partial_tV_G+s\bigl(\mu_{\mathbb Q}(t)+\lambda_I i\bigr)(V_G)_s
        -\kappa_I i(V_G)_i+(\log s)(V_G)_z
    \nonumber\\
    &+\frac12\sigma^2s^2(V_G)_{ss}
       +\frac12\eta(t)^2(V_G)_{ii}
       +\rho\sigma s\eta(t)(V_G)_{si}-rV_G,
    \label{eq:geometric-PDE}
\end{align}
with
\begin{equation}
    V_G(T,s,i,z)=(e^{z/T}-K)^+.
    \label{eq:geometric-terminal}
\end{equation}
The deterministic coefficients under $\mathbb Q$ permit an explicit
calculation of this expectation.

\begin{theorem}[Geometric Asian option price]
\label{thm:geometric-closed-form}
For $K>0$ and $t<T$, the remaining time is $\tau=T-t$. The kernel is
\begin{equation}
    K_I(u)=\int_u^T(T-v)e^{-\kappa_I(v-u)}dv
          =\frac{T-u}{\kappa_I}
            -\frac{1-e^{-\kappa_I(T-u)}}{\kappa_I^2}.
    \label{eq:KI}
\end{equation}
Under $\mathbb Q$, conditional on $(S_t,I_t,Z_t)=(s,i,z)$, $Z_T$ is
Gaussian with mean
\begin{align}
    m_Z(t,s,i,z)
    ={}&z+\tau\log s
       +\int_t^T(T-u)\left(\mu_{\mathbb Q}(u)-\frac12\sigma^2\right)du
    \nonumber\\
    &+\lambda_I iK_I(t),
    \label{eq:mZ}
\end{align}
and variance
\begin{align}
    v_Z(t)
    ={}&\sigma^2\frac{\tau^3}{3}
       +\lambda_I^2\int_t^TK_I(u)^2\eta(u)^2du
    \nonumber\\
    &+2\rho\sigma\lambda_I\int_t^T(T-u)K_I(u)\eta(u)du.
    \label{eq:vZ}
\end{align}
With $\mu_G=m_Z(t,s,i,z)/T$ and $\sigma_G^2=v_Z(t)/T^2$, and,
\begin{equation}
    d_1=\frac{\mu_G-\log K+\sigma_G^2}{\sigma_G},
    \qquad d_2=d_1-\sigma_G.
    \label{eq:d1d2}
\end{equation}
Then the price in \eqref{eq:VG-definition} is
\begin{equation}
    V_G(t,s,i,z)
    =e^{-r(T-t)}\left[
      e^{\mu_G+\frac12\sigma_G^2}N(d_1)-K N(d_2)\right],
    \label{eq:geometric-closed-form}
\end{equation}
where $N$ is the standard normal distribution function. Under
\eqref{eq:cps-spread}, this is an arbitrage-free price in the sense of
Proposition \ref{prop:asian-no-arbitrage}.
\end{theorem}

\begin{proof}
Conditional on $(S_t,I_t,Z_t)=(s,i,z)$, It\^{o}'s formula for
$X_u=\log S_u$ and the solution of \eqref{eq:I-Q} imply
\begin{align}
    dX_u&=\left(\mu_{\mathbb Q}(u)-\frac12\sigma^2
                   +\lambda_I I_u\right)du+\sigma dW_u^{\mathbb Q},
    \label{eq:X-Q}\\
    I_u&=i e^{-\kappa_I(u-t)}
       +\int_t^u e^{-\kappa_I(u-v)}\eta(v)dW_v^{I,\mathbb Q}.
    \label{eq:I-explicit}
\end{align}
By Fubini's theorem,
\begin{align}
    \int_t^T X_u du
    ={}&\tau\log s
       +\int_t^T(T-u)\left(\mu_{\mathbb Q}(u)-\frac12\sigma^2\right)du
    \nonumber\\
       &+\lambda_I\int_t^T(T-u)I_u du
       +\sigma\int_t^T(T-u)dW_u^{\mathbb Q},
    \label{eq:intX-geometric}
\end{align}
and
\[
    \int_t^T(T-u)I_u du
    =iK_I(t)+\int_t^T K_I(u)\eta(u)dW_u^{I,\mathbb Q}.
\]
Both interchanges are valid since the kernels are deterministic and
bounded and the integration domains are finite. We obtain
\begin{equation}
    Z_T=m_Z(t,s,i,z)
       +\sigma\int_t^T(T-u)dW_u^{\mathbb Q}
       +\lambda_I\int_t^T K_I(u)\eta(u)dW_u^{I,\mathbb Q}.
    \label{eq:ZT-Gaussian}
\end{equation}
The two It\^{o} integrals are jointly centered Gaussian. By It\^{o}
isometry and \eqref{eq:corr-Q},
\begin{align*}
    \operatorname{Var}^{\mathbb Q}_{t,s,i,z}(Z_T)
    & =\int_t^T\left[
       \bigl(\sigma(T-u)+\rho\lambda_I K_I(u)\eta(u)\bigr)^2
       +(1-\rho^2)\lambda_I^2K_I(u)^2\eta(u)^2\right]du\\
    &=v_Z(t).
\end{align*}
Expanding the integrand, we obtain \eqref{eq:vZ}.
Since $0\leq K_I(u)\leq(T-u)^2/2$ and $\eta$ is bounded,
\[
    \sigma(T-u)+\rho\lambda_I K_I(u)\eta(u)
       \geq\frac{\sigma}{2}(T-u)>0
\]
for $u<T$ sufficiently close to $T$. Thus $v_Z(t)>0$ for $t<T$.

Consequently, $\log G_T=Z_T/T\sim\mathcal N(\mu_G,\sigma_G^2)$
under the conditional law. Completing the square in the normal density,
\begin{align*}
    \mathbb E^{\mathbb Q}_{t,s,i,z}[(G_T-K)^+]
    &=\frac{1}{\sqrt{2\pi}}\int_{-d_2}^{\infty}
       \left(e^{\mu_G+\sigma_G y}-K\right)e^{-y^2/2}dy\\
    &=e^{\mu_G+\frac12\sigma_G^2}N(d_1)-K N(d_2).
\end{align*}
Multiplication by $e^{-r(T-t)}$ proves
\eqref{eq:geometric-closed-form}. The no-arbitrage assertion follows
from Proposition \ref{prop:asian-no-arbitrage}.
\end{proof}

For the choice $\mu_{\mathbb Q}(u)=r$, the conditional mean simplifies to
\begin{equation}
    m_Z(t,s,i,z)
    =z+\tau\log s+\left(r-\frac12\sigma^2\right)\frac{\tau^2}{2}
      +\lambda_I iK_I(t).
    \label{eq:mZ-r}
\end{equation}
The current value $I_t=i$ contributes $\lambda_I iK_I(t)$ to the
conditional mean of $Z_T$. The term proportional to $\lambda_I^2$ accounts for future random
changes in $I_t$. Their correlation with midpoint shocks contributes
the covariance term. Both the initial order flow state and this
correlation therefore affect the option price.

When $\lambda_I=0$ and $\mu_{\mathbb Q}=r$, the midpoint follows
\[
    dS_t=rS_tdt+\sigma S_tdW_t^{\mathbb Q}.
\]
Subject to the stated compatibility condition when $|\rho|=1$,
the conditional moments reduce to
\[
    m_Z=z+\tau\log s+
         \left(r-\frac12\sigma^2\right)\frac{\tau^2}{2},
    \qquad v_Z=\sigma^2\frac{\tau^3}{3}.
\]
The geometric Asian option formula therefore reduces to the
continuously monitored Kemna--Vorst price \citep{KemnaVorst1990}.

\section{Endogenous valuation}
\label{sec:endogenous}

The dealer's reservation prices account for hedging costs, trading
constraints and risk aversion. We formulate these prices through
exponential utility of terminal wealth under $\mathbb P$, following
\citet{GueantPu2017} and \citet{EkrenNadtochiy2022}.
The option is held until maturity. Purchases and sales of stock are
financed by the cash account at execution prices affected by the
dealer's trading, with market dynamics independent of its strategy.

We specify the trading problem and reservation prices, derive the
formal control equation, and prove a no-arbitrage result for the
dealer's stock trades.

\subsection{The dealer's problem}
\label{subsec:dealer-problem}

For a given initial cash balance, stock holding and option position,
the dealer chooses a trading rate subject to execution and inventory
constraints. Terminal wealth includes the option payment and the
proceeds and costs of liquidating the remaining stock position.

The stock holding is $Q_t$, and its rate of change is $\nu_t$.
Purchases have $\nu_t>0$, and sales have $\nu_t<0$.
The state $J_t$ tracks the transient effect of past trades:
\begin{equation}
    dQ_t=\nu_t dt,\qquad
    dJ_t=(-\kappa_JJ_t+\nu_t)dt,\qquad \kappa_J>0.
    \label{eq:dealer-states}
\end{equation}
The execution price per share is
\begin{equation}
    \widehat S_t=S_t+\bar\lambda_PQ_t+\bar\lambda_TJ_t.
    \label{eq:dealer-execution-price}
\end{equation}
The coefficients satisfy $\bar\lambda_P,\bar\lambda_T\geq0$.
The inventory in this adjustment includes the initial holding. Thus
$q$ shares held at the valuation date contribute $\bar\lambda_Pq$
to the dealer's execution price.
If the dealer stops trading, $Q_t$ stays constant and $J_t$ decays
at rate $\kappa_J$.
The execution price may become negative when the midpoint is small
and the inventory or transient adjustment is negative.
For a fixed $\varepsilon>0$, the dealer trades only when
$\widehat S_t\geq\varepsilon$. When this condition fails, the dealer cannot trade.

Temporary execution costs are paid at the time of a trade and have
no effect on subsequent execution prices. Power law execution costs are used in
\citet{AlmgrenThumHauptmannLi2005} and \citet{GueantPu2017}.
We allow different coefficients for purchases and sales:
\begin{equation}
    C(\nu)=k_A(\nu^+)^{1+\psi}+k_B(\nu^-)^{1+\psi},
    \qquad k_A,k_B>0,
    \label{eq:dealer-temporary-cost}
\end{equation}
where $\nu^+=\max\{\nu,0\}$ and $\nu^-=\max\{-\nu,0\}$.
The cost per share is proportional to $|\nu|^\psi$. We choose the
same exponent $\psi$ as in Section \ref{sec:model}.
The net cash outflow per unit time is
\begin{equation}
    g(s,q,j,\nu)=(s+\bar\lambda_Pq+\bar\lambda_Tj)\nu+C(\nu).
    \label{eq:dealer-cash-rate}
\end{equation}

The dealer's cash account $X_t$ earns interest at the risk-free rate
$r\geq0$ and pays for trades:
\begin{equation}
    dX_t=rX_tdt-g(S_t,Q_t,J_t,\nu_t)dt.
    \label{eq:dealer-cash}
\end{equation}
The mark-to-market wealth satisfies
\begin{align}
    d(X_t+Q_tS_t)
    ={}&rX_tdt+Q_tdS_t
    \nonumber\\
       &-\bigl[(\bar\lambda_PQ_t+\bar\lambda_TJ_t)\nu_t
                     +C(\nu_t)\bigr]dt.
    \label{eq:dealer-wealth}
\end{align}

The dealer's problem uses a weak formulation. An admissible system
consists of a filtered probability space, the Brownian motions
$(W_u,W_u^I)_{t\leq u\leq T}$ with correlation $\rho$, and a
progressively measurable trading rate. 

Starting from $(t,s,i,q,j,a)$ with $s>0$ and $|q|\leq\bar Q$,
admissible trading rates satisfy,
for almost every $u\in[t,T]$, a.s.,
\begin{equation}
    |\nu_u|\leq\bar\nu,\qquad |Q_u|\leq\bar Q,
    \qquad \nu_u=0\quad\text{if }\widehat S_u<\varepsilon,
    \label{eq:dealer-constraints}
\end{equation}
where $\bar\nu,\bar Q>0$. They form the admissible class
$\mathcal A(t,s,i,q,j,a)$. We index a system by its trading rate
$\nu$ and use $\mathbb P$ for its probability measure. Each supremum
over $\mathcal A$ includes the choice of that system.
The cap $\bar\nu$ limits execution capacity to $\bar\nu(u-t)$
shares over $[t,u]$, ruling out an immediate change in the hedge.
It also ensures $\int_t^T C(\nu_u)du<\infty$. The bound $\bar Q$
limits the stock position.

At expiry the remaining stock is liquidated at $S_T$, with a charge
$L(Q_T)$. The rate limit applies to trading before expiry and exempts
this terminal transaction. A terminal liquidation cost also appears in
option hedging with execution frictions \citep{GueantPu2017}.
We use
\begin{equation}
    L(q)=\ell_1|q|+\frac{\Gamma_Q}{2}q^2,
    \qquad \ell_1,\Gamma_Q\geq0.
    \label{eq:dealer-liquidation}
\end{equation}
The charge has a per-share component, for a spread or fee, and a
quadratic component for the cost of liquidating a larger position.
We impose no further charge on $J_T$.

The option position has size $n>0$.
We use $\theta=+1$ for a short option position, $\theta=-1$ for a long
position and $\theta=0$ when no option is held.
Terminal wealth is
\begin{equation}
    W_T^{\theta,\nu}
    =X_T+Q_TS_T-L(Q_T)-\theta n\Phi(R_T).
    \label{eq:dealer-terminal-wealth}
\end{equation}
The payoff $\Phi$ is the arithmetic or geometric Asian call option payoff
defined in Section \ref{sec:model}.

For exponential utility $U(w)=-e^{-\gamma w}$, with risk aversion
parameter $\gamma>0$, the utility value for position
$\theta$ is
\begin{equation}
    u^\theta(t,x,s,i,q,j,a)
    =\sup_{\nu\in\mathcal A(t,s,i,q,j,a)}
       \mathbb E^{\mathbb P}_{t,x,s,i,q,j,a}
          [-e^{-\gamma W_T^{\theta,\nu}}].
    \label{eq:dealer-utility-value}
\end{equation}
When the relevant utility values are finite, the reservation ask and
bid are defined by
\begin{align}
    u^{+1}(t,x+P_{\rm ask},s,i,q,j,a)&=u^0(t,x,s,i,q,j,a),
    \label{eq:dealer-ask-definition}\\
    u^{-1}(t,x-P_{\rm bid},s,i,q,j,a)&=u^0(t,x,s,i,q,j,a).
    \label{eq:dealer-bid-definition}
\end{align}
At time $t$, the seller receives $P_{\rm ask}$ and the buyer pays
$P_{\rm bid}$. Each payment equates the utility value with that of
holding no option.

For $u^\theta>-\infty$, there must be at least one admissible
strategy $\nu$ satisfying
\begin{equation}
    \mathbb E^{\mathbb P}
      \left[e^{\gamma(W_T^{\theta,\nu})^-}\right]<\infty,
    \label{eq:dealer-exponential-integrability}
\end{equation}
where $y^-=\max\{-y,0\}$.
This condition is required for $\theta=0,+1$ when defining the
reservation ask, and for $\theta=0,-1$ when defining the reservation bid.
A strategy with infinite expected exponential loss has utility $-\infty$. When $\lambda_I=0$ and $Q_0=J_0=R_0=0$, the integrability
condition \eqref{eq:dealer-exponential-integrability} fails
for every admissible seller strategy by
Proposition \ref{prop:dealer-zero-inventory}. When the condition holds,
$u^\theta>-\infty$. The bounds on trading and inventory also imply
$u^\theta<0$ by Jensen's inequality, so $\log(-u^\theta)$ is well
defined.

Exponential utility allows us to factor out initial cash and define
$v^\theta$, the value function used in the control equation
\citep{henderson2004utility,GueantPu2017}:
\begin{equation}
    u^\theta(t,x,s,i,q,j,a)
    =-\exp\{-\gamma[e^{r(T-t)}x+v^\theta(t,s,i,q,j,a)]\},
    \label{eq:dealer-cara}
\end{equation}
The terminal condition is
\begin{equation}
    v^\theta(T,s,i,q,j,a)=qs-L(q)-\theta n\Phi(a).
    \label{eq:dealer-terminal-v}
\end{equation}
Substituting \eqref{eq:dealer-cara} into the definitions
\eqref{eq:dealer-ask-definition}--\eqref{eq:dealer-bid-definition},
we obtain
\begin{equation}
    P_{\rm ask}=e^{-r(T-t)}(v^0-v^{+1}),\qquad
    P_{\rm bid}=e^{-r(T-t)}(v^{-1}-v^0).
    \label{eq:dealer-reservation-prices}
\end{equation}
Both prices refer to $n$ options and are calculated using the same
initial cash, stock holding and transient impact state.

For either payoff, a sufficient condition for finite reservation
bid and ask prices is
\[
    \frac nT\int_t^T e^{-r(T-u)}du\leq q\leq\bar Q,
    \qquad \bar\nu\geq n/T,
\]
at a finite initial state with $s>0$ and $a=0$ when $t=0$
(Proposition \ref{prop:dealer-covered}).

\subsection{The control equation}
\label{subsec:dealer-hjb}

We derive the formal Hamilton--Jacobi--Bellman equation for
$v^\theta$. For the full state
$\mathbf Z_u^\nu=(X_u,S_u,I_u,Q_u,J_u,R_u)$, initially
$\mathbf z=(x,s,i,q,j,a)$, the dynamic programming identity is
\begin{equation}
 u^\theta(t,\mathbf z)
 =\sup_{\nu\in\mathcal A(t,s,i,q,j,a)}
    \mathbb E^{\mathbb P}[u^\theta(\tau,\mathbf Z_\tau^\nu)],
 \label{eq:dealer-dpp}
\end{equation}
where $t\leq\tau\leq T$ is a stopping rule based on the Brownian,
market and trading history, evaluated in each admissible system.
Expectations may equal $-\infty$. We apply
\citet[Theorem 3.4]{ElKarouiTan2024}.\footnote{We use $\log S_u$ as the price coordinate and adjoin the
Brownian coordinates and
\[
 H_u=\int_t^u\left(\mathbf 1_{\{|Q_v|>\bar Q\}}
       +|\nu_v|\mathbf 1_{\{\widehat S_v<\varepsilon\}}\right)dv.
\]
With rates in $[-\bar\nu,\bar\nu]$, $H_T=0$ imposes the inventory
and execution restrictions. The state is frozen after $T$. The
reward is terminal utility for $H_T=0$ and termination at or after
$T$, and $-\infty$ otherwise. The coefficients are Borel measurable
and satisfy the theorem's integrability condition. The reward is
Borel measurable and bounded above by zero. Along admissible
histories, $H_u=0$ and the continuation value depends only on
$\mathbf Z_u^\nu$.}

For a smooth test function $f=f(t,s,i,q,j,a)$, the generator of the
external states is
\begin{align}
 \mathcal L^{\mathbb P}f
 ={}&s(\mu(t)+\lambda_Ii)f_s-\kappa_Iif_i+h(s)f_a
 \nonumber\\
 &+\frac12\sigma^2s^2f_{ss}+\frac12\eta(t)^2f_{ii}
   +\rho\sigma s\eta(t)f_{si}.
 \label{eq:dealer-physical-generator}
\end{align}
The inventory bounds restrict the trading rates to
\begin{equation}
 \mathcal U(q)=
 \begin{cases}
 [-\bar\nu,\bar\nu],&|q|<\bar Q,\\
 [-\bar\nu,0],&q=\bar Q,\\
 [0,\bar\nu],&q=-\bar Q.
 \end{cases}
 \label{eq:dealer-rate-interval}
\end{equation}
The execution-price condition further requires
\[
 \mathcal U(s,q,j)=
 \begin{cases}
 \mathcal U(q),&s+\bar\lambda_Pq+\bar\lambda_Tj\geq\varepsilon,\\
 \{0\},&s+\bar\lambda_Pq+\bar\lambda_Tj<\varepsilon.
 \end{cases}
\]
The instantaneous variance of the diffusion term in $f$ is
\begin{equation}
 \mathcal R(f)=\sigma^2s^2f_s^2
       +2\rho\sigma s\eta(t)f_sf_i+\eta(t)^2f_i^2.
 \label{eq:dealer-risk-term}
\end{equation}
A formal application of dynamic programming and It\^{o}'s formula
to \eqref{eq:dealer-cara}, assuming the required derivatives exist,
leads to the reduced control equation
\citep[Section 3.4.1]{pham2009continuous}:
\begin{align}
 0={}&v_t^\theta+\mathcal L^{\mathbb P}v^\theta
       -\kappa_Jjv_j^\theta-\frac\gamma2\mathcal R(v^\theta)
 \nonumber\\
 &+\max_{\nu\in\mathcal U(s,q,j)}
   \{\nu(v_q^\theta+v_j^\theta)-e^{r(T-t)}g(s,q,j,\nu)\},
 \label{eq:dealer-reduced-hjb}
\end{align}
with terminal value \eqref{eq:dealer-terminal-v}.

Equation \eqref{eq:dealer-reduced-hjb} follows formally from dynamic
programming where the value function has the required derivatives,
at interior states away from the execution-price threshold.
Section \ref{subsec:num-correctness} compares numerical solutions
of this equation with Bellman estimates of the utility-based
reservation prices.

\subsection{No arbitrage}
\label{subsec:dealer-price-comparisons}

The execution and terminal liquidation rules must exclude profits
generated solely by the dealer's own impact. This is the price
manipulation issue studied by \citet{HubermanStanzl2004} and
\citet{Gatheral2010}. For the present model, we prove that the
cumulative impact cost is nonnegative when $Q_0=J_0=0$, including
for round trips with $Q_T=0$.

We then rule out arbitrage from stock trading with zero initial cash,
inventory and transient impact. The proof uses an equivalent measure
$\mathbb Q^0$ under which the discounted market price is a martingale.
Discounted trading gains have zero expectation under this measure,
and the nonnegative costs imply nonpositive expected discounted
terminal wealth.
The result concerns the dealer's stock trades. Utility is evaluated
under $\mathbb P$, and the measure $\mathbb Q$ in
Section \ref{sec:exogenous} determines the exogenous Asian option price.
We also use $\mathbb Q^0$ for the frictionless comparison below.

On each admissible probability space, the market price of risk and
density are
\[
 \phi_t^0=\frac{\mu(t)+\lambda_I I_t-r}{\sigma},\qquad
 Z_t^0=\exp\left\{-\int_0^t\phi_u^0dW_u
                     -\frac12\int_0^t(\phi_u^0)^2du\right\}.
\]
For $\phi(t,i)=(\mu(t)+\lambda_Ii-r)/\sigma$, the drifts
$-\kappa_Ii$ and $-\kappa_Ii-\rho\eta(t)\phi(t,i)$ and the
coefficient $\rho\phi(t,i)$ satisfy linear growth bounds, while
$\eta$ is bounded. By
\citet[Theorem 4.1]{KlebanerLiptser2014},
\[
 Z_t^I=\exp\left\{-\rho\int_0^t\phi_u^0dW_u^I
       -\frac{\rho^2}{2}\int_0^t(\phi_u^0)^2du\right\},
 \qquad \mathbb E^{\mathbb P}[Z_T^I]=1.
\]
For $|\rho|<1$, the decomposition
$W_t=\rho W_t^I+\sqrt{1-\rho^2}W_t^\perp$ has $W_t^\perp$
independent of $W_t^I$. Conditional on $(W_u^I)_{0\leq u\leq T}$,
the exponential factor $Z_T^0/Z_T^I$ has a deterministic
$L^2([0,T])$ integrand, so
$\mathbb E^{\mathbb P}[Z_T^0/Z_T^I\mid (W_u^I)_{0\leq u\leq T}]=1$.
Hence $\mathbb E^{\mathbb P}[Z_T^0]=1$. For $|\rho|=1$,
$Z_T^0=Z_T^I$. Since $Z_T^0>0$ a.s.,
$d\mathbb Q^0=Z_T^0d\mathbb P$ defines an equivalent measure.
From Girsanov's theorem,
\[
 dS_t=rS_tdt+\sigma S_tdW_t^0,\qquad
 \mathbb E^{\mathbb Q^0}[(e^{-rt}S_t)^2]=S_0^2e^{\sigma^2t},
\]
so $(e^{-rt}S_t)_{0\leq t\leq T}$ is an
$L^2(\mathbb Q^0)$ martingale.

\begin{prop}
\label{prop:dealer-no-arbitrage}
Suppose $r\geq0$, $\bar\lambda_P,\bar\lambda_T\geq0$, and $C,L\geq0$.
For $X_0=Q_0=J_0=0$ and no option endowment, every admissible
strategy satisfies
\[
 \mathbb E^{\mathbb Q^0}
   [e^{-rT}(X_T+Q_TS_T-L(Q_T))]\leq0.
\]
Hence stock trading admits no arbitrage, and round trips with $Q_T=0$
have nonpositive expected discounted cash gain under $\mathbb Q^0$.

\end{prop}

\begin{proof}
The discounted midpoint is $\bar S_t=e^{-rt}S_t$.
Discounting the wealth equation \eqref{eq:dealer-wealth}, integrating
from $0$ to $T$, and deducting the terminal liquidation cost, we obtain
\[
 e^{-rT}(X_T+Q_TS_T-L(Q_T))
 =\int_0^TQ_u d\bar S_u-K_T-e^{-rT}L(Q_T),
\]
where the discounted impact and temporary execution costs are
\[
 K_T=\int_0^T e^{-ru}
   [(\bar\lambda_PQ_u+\bar\lambda_TJ_u)\nu_u+C(\nu_u)]du.
\]
Since $Q_0=J_0=0$, the identities
\eqref{eq:dealer-permanent-cost} and \eqref{eq:dealer-transient-cost}
imply $K_T\geq0$. Since the discounted midpoint is an
$L^2(\mathbb Q^0)$ martingale and $|Q_u|\leq\bar Q$,
\[
 \mathbb E^{\mathbb Q^0}\left[\int_0^TQ_u d\bar S_u\right]=0.
\]
The cost terms are integrable under the rate and inventory bounds.
Taking expectations and using $L\geq0$ proves the stated inequality.
If $W_T^{0,\nu}\geq0$ a.s. and
$\mathbb P(W_T^{0,\nu}>0)>0$, equivalence implies
$\mathbb E^{\mathbb Q^0}[e^{-rT}W_T^{0,\nu}]>0$,
contradicting the inequality.
For a round trip, $Q_T=0$ and $L(0)=0$, so the terminal gain is $X_T$.

\end{proof}

The expectation inequality is under $\mathbb Q^0$.
Expected trading profits under $\mathbb P$ may be positive.

Setting $\bar\lambda_P=\bar\lambda_T=0$ removes the dealer's
inventory and transient adjustments to execution prices. Temporary
costs, terminal liquidation costs and trading restrictions still
affect the hedge and can produce different reservation bid and ask
prices. In the fully frictionless case, $C=L=0$ and the trading rate,
inventory and execution-price restrictions are also removed.
The unit-claim replication price is
\[
 F(t,s,a)=\mathbb E^{\mathbb Q^0}_{t,s,a}
          [e^{-r(T-t)}\Phi(R_T)].
\]
If $-\infty<u^0<0$ and admissible strategies permit addition and
subtraction of the replicating hedge, then
\begin{equation}
 P_{\rm bid}^{\rm fr}=P_{\rm ask}^{\rm fr}=nF(t,s,a).
 \label{eq:dealer-frictionless-prices}
\end{equation}
This is the replicable-claim property of utility indifference prices
\citep[Section 2.3, property (ii)]{henderson2004utility}.
Under $\mathbb Q^0$, the discounted Asian payoff belongs to
$L^2(\mathbb Q^0)$ and depends only on the stock price path.
It therefore admits a self-financing hedge in the stock and
bank account
\citep[Theorem 12.4]
{bjork2009arbitrage}.

For the geometric Asian call option, $F$ is the continuously
monitored Kemna--Vorst price. For the arithmetic Asian call
option, $F$ is the corresponding frictionless arithmetic
value. Under the assumptions of
Section \ref{sec:exogenous}, with $\lambda_I=0$ and
$\mu_{\mathbb Q}=r$, its exogenous value also equals $F$
for the same strike, maturity and averaging convention.

For the linear utility objective $U(w)=w$, at
$Q_0=J_0=0$, $\lambda_I=0$ and $\mu=r$, the bid and ask also
equal $nF$. Stock gains have zero expected discounted value and
execution and liquidation costs are nonnegative, so $\nu_t=0$
is optimal. Under the assumptions of
Proposition \ref{prop:dealer-zero-inventory}, the seller's value
function satisfies $u^{+1}=-\infty$, and its reservation ask is
$P_{\rm ask}=+\infty$ for every $\gamma>0$.

\begingroup
\renewcommand{\topfraction}{0.95}
\renewcommand{\bottomfraction}{0.8}
\renewcommand{\textfraction}{0.05}
\renewcommand{\floatpagefraction}{0.85}
\setcounter{topnumber}{3}
\setcounter{totalnumber}{4}
\setlength{\textfloatsep}{10pt plus 2pt minus 4pt}
\setlength{\floatsep}{8pt plus 2pt minus 2pt}
\suppressfloats[t]
\section{Numerical results}
\label{sec:numerics}

We calculate the effect of external order flow on exogenous prices
and the effect of hedging costs and risk on dealer quotes. After
specifying the Bellman calculation, we examine exogenous prices,
compare implementations of the dealer's valuation and study the
resulting spreads and stock trades.

\subsection{Bellman scheme and numerical setup}
\label{subsec:num-setup}
\label{subsec:dealer-bellman}

The dealer's reservation prices are computed by backward induction
for $\theta\in\{-1,0,+1\}$. 
The trading dates are $t_m=m\Delta t$, with $\Delta t=T/N$.
As in Section \ref{sec:model}, the four pairs
$(\xi,\zeta)\in\{-1,1\}^2$ have probabilities
$p_{\xi,\zeta}=(1+\rho\xi\zeta)/4$.
From state $(s,i,q,j,a)$ at $t_m$, a trading rate $\nu$
produces the following state at $t_{m+1}$:

\begin{align}
    s'_\xi&=s\exp\left\{
       (\mu(t_m)+\lambda_Ii-\tfrac12\sigma^2)\Delta t
                   +\sigma\sqrt{\Delta t}\xi\right\},
    \label{eq:dealer-step-S}\\
    i'_\zeta&=i-\kappa_Ii\Delta t+\eta(t_m)\sqrt{\Delta t}\zeta,
    \label{eq:dealer-step-I}\\
    q'&=q+\nu\Delta t,\qquad
    j'=j+(-\kappa_Jj+\nu)\Delta t,\qquad
    a'_\xi=a+\tfrac12[h(s)+h(s'_\xi)]\Delta t.
    \label{eq:dealer-step-other}
\end{align}
We use the trapezoidal rule to calculate $R_T$, assigning half
the weight of an interior observation to each endpoint.
Using $g(s,q,j,\nu)$ as the cash outflow per unit time over
$[t_m,t_{m+1}]$, the cash balance becomes
\begin{equation}
    x'=e^{r\Delta t}\bigl[x-\delta_r g(s,q,j,\nu)\bigr],\qquad
    \delta_r=
    \begin{cases}
      (1-e^{-r\Delta t})/r,&r\ne0,\\
      \Delta t,&r=0.
    \end{cases}
    \label{eq:dealer-step-cash}
\end{equation}
For a finite candidate set
$\mathcal U_N\subset[-\bar\nu,\bar\nu]$ containing zero, the
available rates at $(s,q,j)$ are
\begin{equation}
    \mathcal U_m(s,q,j)=
    \begin{cases}
      \{\nu\in\mathcal U_N:|q+\nu\Delta t|\leq\bar Q\},
        &s+\bar\lambda_Pq+\bar\lambda_Tj\geq\varepsilon,\\
      \{0\},&s+\bar\lambda_Pq+\bar\lambda_Tj<\varepsilon.
    \end{cases}
    \label{eq:dealer-discrete-rates}
\end{equation}
The execution-price condition is checked at $t_m$.
The discrete value functions $v_m^\theta$ have terminal values
\[
    v_N^\theta(s,i,q,j,a)=qs-L(q)-\theta n\Phi(a),
    \qquad \theta\in\{-1,0,+1\}.
\]
For $m=N-1,\ldots,0$, the Bellman update is
\begin{equation}
\begin{split}
    v_m^\theta(s,i,q,j,a)
    ={}&\max_{\nu\in\mathcal U_m(s,q,j)}
       \Bigg\{-e^{r(T-t_m)}\delta_r g(s,q,j,\nu)\\
       &\qquad-\frac1\gamma\log\Bigg[
          \sum_{\xi,\zeta}p_{\xi,\zeta}
          e^{-\gamma\mathcal I v_{m+1}^\theta
                   (s'_\xi,i'_\zeta,q',j',a'_\xi)}\Bigg]\Bigg\}.
\end{split}
    \label{eq:dealer-bellman}
\end{equation}
Here $\mathcal I$ denotes multilinear interpolation of continuation
values. The bid and ask are calculated from the three initial values
using \eqref{eq:dealer-reservation-prices}.

The Bellman calculations produce reservation prices for the finite
step problem. At zero initial inventory, under the conditions of
Proposition \ref{prop:dealer-zero-inventory}, the corresponding
continuous time ask is infinite. The comparisons below also include
positive initial inventories satisfying the sufficient condition for
finite continuous time reservation prices in
Proposition \ref{prop:dealer-covered}.

The numerical experiments use one option with $S_0=K=100$ and $T=1$.
The baseline market parameters are
\[
 \sigma=0.20,\quad r=\mu=\mu_{\mathbb Q}=0.05,\quad
 \lambda_I=0.05,\quad \kappa_I=1,\quad \eta=0.5,\quad
 \rho=I_0=0.
\]
The dealer's parameters are
\[
 \bar\lambda_T=0.05,\quad \bar\lambda_P=0.025,\quad
 \kappa_J=1,\quad k_A=k_B=0.5,\quad \psi=1,\quad \gamma=0.05,
\]
\[
 \Gamma_Q=1,\quad \ell_1=0,\quad \bar Q=3,\quad \bar\nu=5,
 \quad Q_0=J_0=0,\quad n=1.
\]
The dealer calculations in Tables \ref{tab:num-correctness} and
\ref{tab:spread-sens} use $\lambda_I=0$, as do Panels A--D of
Table \ref{tab:three-model}. Panel E varies $\lambda_I$ in both
valuations. Other parameter changes are stated with each comparison.

The baseline calculation uses $15$ equally spaced trading rates in
$[-5,5]$, subject to the inventory and execution-price restrictions.
The baseline uses $N=25$ trading intervals and $61$, $15$, $121$
and $9$ nodes for $Q_t$, $J_t$, $R_t$ and $I_t$, respectively.

For Table \ref{tab:num-correctness}, all cases use
$\lambda_I=\rho=I_0=J_0=0$, with the other costs and constraints at
baseline values. The inventories $Q_0=1,2$ satisfy the finiteness
condition in Proposition \ref{prop:dealer-covered}; at $Q_0=0$, only
the bid is compared. Bellman estimates are extrapolated from
$N=25,50,100$, using $\Delta\nu=1,0.5$, $\Delta Q=\Delta\nu/N$ and
$n_R=121,241,481,961$, as available at each $N$. The target is
$n_J=5$, with $n_J=3$ in some calculations for Panels B and D.
After corrections for $R_t$ and rate spacings, prices and spreads are
each extrapolated quadratically in $1/N$. At $N=100$, the $R_t$
correction is scaled from coarser calculations and the rate correction
is borrowed from $N=50$. The Policy estimates use trading rules from
one GPI run, evaluated on $10^6$ simulated paths with $400$ trading dates.

Geometric exogenous prices are computed from
Theorem \ref{thm:geometric-closed-form}, with the Kemna--Vorst price
(KV) as their frictionless reference. Arithmetic exogenous prices
and their frictionless references are estimated using $2\times10^5$
Monte Carlo paths and the equally weighted average of $253$ prices,
including $S_0$ and $S_T$.

\subsection{Exogenous valuation}
\label{subsec:num-exogenous}

Table \ref{tab:exog-geom} and Figure \ref{fig:num-exog} report
arithmetic and geometric option prices as external order flow varies.
For each payoff we compare the price with its frictionless reference:
the Kemna--Vorst formula for geometric averaging and a Monte Carlo
estimate for arithmetic averaging.

\begin{table}[tbp]
\centering
\scriptsize
\setlength{\tabcolsep}{4pt}
\begin{tabular}{@{}lrrrr|rrr@{}}
\hline
 & & \multicolumn{3}{c|}{\textbf{Geometric option}}
   & \multicolumn{3}{c}{\textbf{Arithmetic option}} \\
\cline{3-5}\cline{6-8}
\textbf{Par.} & \textbf{Value}
  & \textbf{KV} & \textbf{Price} & \textbf{Imp.\ \%}
  & \textbf{Ref.} & \textbf{Price} & \textbf{Imp.\ \%} \\
\hline
\multicolumn{5}{@{}l|}{\emph{Panel A: drift loading $\lambda_I$}} & \multicolumn{3}{l}{} \\
$\lambda_I$ & 0.000 & 5.5468 & 5.5468 & 0.000 & \shortstack[r]{5.7647\\[-1pt]{\tiny $\pm$0.0015}} & \shortstack[r]{5.7647\\[-1pt]{\tiny $\pm$0.0015}} & 0.000 \\
$\lambda_I$ & 0.025 & 5.5468 & 5.5477 & 0.016 & \shortstack[r]{5.7647\\[-1pt]{\tiny $\pm$0.0015}} & \shortstack[r]{5.7656\\[-1pt]{\tiny $\pm$0.0015}} & 0.017 \\
$\lambda_I$ & 0.050 & 5.5468 & 5.5504 & 0.065 & \shortstack[r]{5.7647\\[-1pt]{\tiny $\pm$0.0015}} & \shortstack[r]{5.7686\\[-1pt]{\tiny $\pm$0.0016}} & 0.068 \\
$\lambda_I$ & 0.075 & 5.5468 & 5.5549 & 0.145 & \shortstack[r]{5.7647\\[-1pt]{\tiny $\pm$0.0015}} & \shortstack[r]{5.7736\\[-1pt]{\tiny $\pm$0.0016}} & 0.154 \\
$\lambda_I$ & 0.100 & 5.5468 & 5.5612 & 0.258 & \shortstack[r]{5.7647\\[-1pt]{\tiny $\pm$0.0015}} & \shortstack[r]{5.7805\\[-1pt]{\tiny $\pm$0.0016}} & 0.275 \\
$\lambda_I$ & 0.125 & 5.5468 & 5.5692 & 0.404 & \shortstack[r]{5.7647\\[-1pt]{\tiny $\pm$0.0015}} & \shortstack[r]{5.7894\\[-1pt]{\tiny $\pm$0.0016}} & 0.430 \\
$\lambda_I$ & 0.150 & 5.5468 & 5.5790 & 0.581 & \shortstack[r]{5.7647\\[-1pt]{\tiny $\pm$0.0015}} & \shortstack[r]{5.8003\\[-1pt]{\tiny $\pm$0.0016}} & 0.618 \\
\multicolumn{5}{@{}l|}{\emph{Panel B: initial impact state $I_0$}} & \multicolumn{3}{l}{} \\
$I_0$ & $-$1.00 & 5.5468 & 4.5479 & $-$18.009 & \shortstack[r]{5.7647\\[-1pt]{\tiny $\pm$0.0015}} & \shortstack[r]{4.7275\\[-1pt]{\tiny $\pm$0.0014}} & $-$17.991 \\
$I_0$ & $-$0.50 & 5.5468 & 5.0330 & $-$9.264 & \shortstack[r]{5.7647\\[-1pt]{\tiny $\pm$0.0015}} & \shortstack[r]{5.2313\\[-1pt]{\tiny $\pm$0.0015}} & $-$9.253 \\
$I_0$ & $-$0.25 & 5.5468 & 5.2876 & $-$4.673 & \shortstack[r]{5.7647\\[-1pt]{\tiny $\pm$0.0015}} & \shortstack[r]{5.4957\\[-1pt]{\tiny $\pm$0.0015}} & $-$4.666 \\
$I_0$ & 0.00 & 5.5468 & 5.5504 & 0.065 & \shortstack[r]{5.7647\\[-1pt]{\tiny $\pm$0.0015}} & \shortstack[r]{5.7686\\[-1pt]{\tiny $\pm$0.0016}} & 0.068 \\
$I_0$ & 0.25 & 5.5468 & 5.8213 & 4.948 & \shortstack[r]{5.7647\\[-1pt]{\tiny $\pm$0.0015}} & \shortstack[r]{6.0500\\[-1pt]{\tiny $\pm$0.0016}} & 4.949 \\
$I_0$ & 0.50 & 5.5468 & 6.1003 & 9.978 & \shortstack[r]{5.7647\\[-1pt]{\tiny $\pm$0.0015}} & \shortstack[r]{6.3398\\[-1pt]{\tiny $\pm$0.0016}} & 9.977 \\
$I_0$ & 1.00 & 5.5468 & 6.6825 & 20.475 & \shortstack[r]{5.7647\\[-1pt]{\tiny $\pm$0.0015}} & \shortstack[r]{6.9449\\[-1pt]{\tiny $\pm$0.0017}} & 20.474 \\
\multicolumn{5}{@{}l|}{\emph{Panel C: mean reversion $\kappa_I$}} & \multicolumn{3}{l}{} \\
$\kappa_I$ & 0.25 & 5.5468 & 5.5521 & 0.094 & \shortstack[r]{5.7647\\[-1pt]{\tiny $\pm$0.0015}} & \shortstack[r]{5.7705\\[-1pt]{\tiny $\pm$0.0016}} & 0.100 \\
$\kappa_I$ & 0.50 & 5.5468 & 5.5514 & 0.083 & \shortstack[r]{5.7647\\[-1pt]{\tiny $\pm$0.0015}} & \shortstack[r]{5.7697\\[-1pt]{\tiny $\pm$0.0016}} & 0.088 \\
$\kappa_I$ & 1.00 & 5.5468 & 5.5504 & 0.065 & \shortstack[r]{5.7647\\[-1pt]{\tiny $\pm$0.0015}} & \shortstack[r]{5.7686\\[-1pt]{\tiny $\pm$0.0016}} & 0.068 \\
$\kappa_I$ & 2.00 & 5.5468 & 5.5491 & 0.042 & \shortstack[r]{5.7647\\[-1pt]{\tiny $\pm$0.0015}} & \shortstack[r]{5.7672\\[-1pt]{\tiny $\pm$0.0015}} & 0.044 \\
$\kappa_I$ & 3.00 & 5.5468 & 5.5484 & 0.028 & \shortstack[r]{5.7647\\[-1pt]{\tiny $\pm$0.0015}} & \shortstack[r]{5.7664\\[-1pt]{\tiny $\pm$0.0015}} & 0.030 \\
$\kappa_I$ & 5.00 & 5.5468 & 5.5477 & 0.015 & \shortstack[r]{5.7647\\[-1pt]{\tiny $\pm$0.0015}} & \shortstack[r]{5.7656\\[-1pt]{\tiny $\pm$0.0015}} & 0.016 \\
$\kappa_I$ & 10.00 & 5.5468 & 5.5471 & 0.005 & \shortstack[r]{5.7647\\[-1pt]{\tiny $\pm$0.0015}} & \shortstack[r]{5.7650\\[-1pt]{\tiny $\pm$0.0015}} & 0.005 \\
\multicolumn{5}{@{}l|}{\emph{Panel D: correlation $\rho$}} & \multicolumn{3}{l}{} \\
$\rho$ & $-$0.75 & 5.5468 & 5.4084 & $-$2.496 & \shortstack[r]{5.7647\\[-1pt]{\tiny $\pm$0.0015}} & \shortstack[r]{5.6175\\[-1pt]{\tiny $\pm$0.0015}} & $-$2.552 \\
$\rho$ & $-$0.50 & 5.5468 & 5.4561 & $-$1.636 & \shortstack[r]{5.7647\\[-1pt]{\tiny $\pm$0.0015}} & \shortstack[r]{5.6683\\[-1pt]{\tiny $\pm$0.0015}} & $-$1.672 \\
$\rho$ & $-$0.25 & 5.5468 & 5.5034 & $-$0.782 & \shortstack[r]{5.7647\\[-1pt]{\tiny $\pm$0.0015}} & \shortstack[r]{5.7186\\[-1pt]{\tiny $\pm$0.0015}} & $-$0.799 \\
$\rho$ & 0.00 & 5.5468 & 5.5504 & 0.065 & \shortstack[r]{5.7647\\[-1pt]{\tiny $\pm$0.0015}} & \shortstack[r]{5.7686\\[-1pt]{\tiny $\pm$0.0016}} & 0.068 \\
$\rho$ & 0.25 & 5.5468 & 5.5970 & 0.905 & \shortstack[r]{5.7647\\[-1pt]{\tiny $\pm$0.0015}} & \shortstack[r]{5.8182\\[-1pt]{\tiny $\pm$0.0016}} & 0.929 \\
$\rho$ & 0.50 & 5.5468 & 5.6433 & 1.739 & \shortstack[r]{5.7647\\[-1pt]{\tiny $\pm$0.0015}} & \shortstack[r]{5.8675\\[-1pt]{\tiny $\pm$0.0016}} & 1.785 \\
$\rho$ & 0.75 & 5.5468 & 5.6892 & 2.568 & \shortstack[r]{5.7647\\[-1pt]{\tiny $\pm$0.0015}} & \shortstack[r]{5.9165\\[-1pt]{\tiny $\pm$0.0016}} & 2.634 \\
\hline
\end{tabular}
\caption{Geometric and arithmetic Asian option prices under $\mathbb Q$,
with $\mu_{\mathbb Q}=r$. KV is the geometric Kemna--Vorst price,
and Ref.\ is the arithmetic Monte Carlo reference. Imp.\ \% is
$100(\mathrm{Price}-P_{\rm ref})/P_{\rm ref}$, using the respective
reference $P_{\rm ref}$. Monte Carlo: nominal $95\%$ sampling
half widths for the arithmetic estimates, without averaging or
control-variate centring errors. In each panel, all parameters except
the indicated one have the baseline values of
Section \ref{subsec:num-setup}.}
\label{tab:exog-geom}
\end{table}

\begin{figure}[tbp]
\centering
\includegraphics[width=\textwidth]{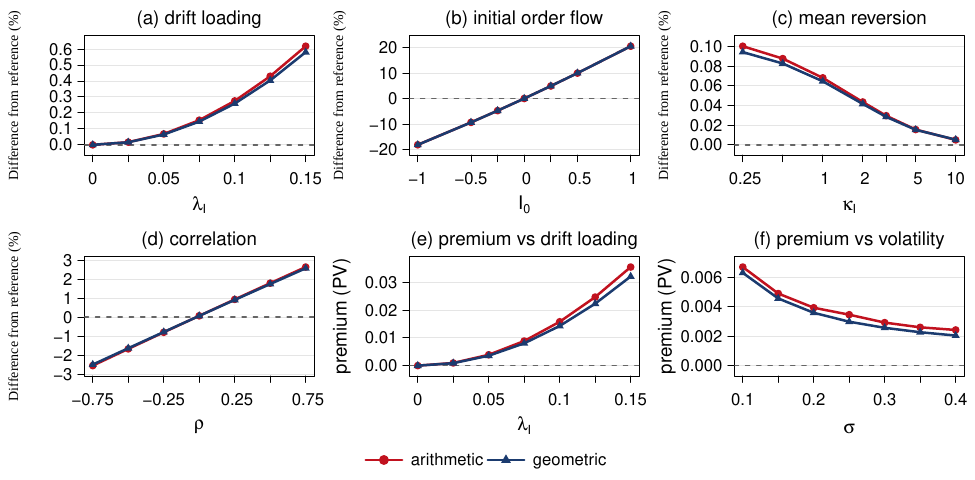}
\caption{Exogenous Asian option prices relative to their frictionless references.
(a)--(d) Percentage differences for the arithmetic and geometric
options in Table \ref{tab:exog-geom}. (e)--(f) Differences in present
value for arithmetic and geometric options as $\lambda_I$ and $\sigma$
vary. Red circles denote arithmetic options and blue triangles
geometric options. The arithmetic estimates use Monte Carlo simulation.}
\label{fig:num-exog}
\end{figure}

Initial order flow has the largest effect in Table \ref{tab:exog-geom}.
In Panel B, both option prices increase substantially with $I_0$, with
nearly equal percentage changes. For $\lambda_I>0$, the return
contribution $\lambda_I I_t$ is initially positive when $I_0>0$ and
negative when $I_0<0$. This contribution decays over time. The prices affected by it remain
part of the average. Arithmetic prices are higher in the
table, as one would expect from the ordering of the two payoffs under
a common averaging convention.

The smaller premiums in Panels A and C arise with $I_0=\rho=0$.
Here random fluctuations of $I_t$ introduce additional variability
in returns. Increasing $\lambda_I$ amplifies their contribution and
raises the call price. Increasing $\kappa_I$ shortens their duration.
The premium therefore falls towards zero as mean reversion becomes
faster over the range in Panel C.

Panel D varies the correlation between price innovations and changes
in $I_t$. At positive $\rho$, a price shock tends to coincide with
an order flow shock that moves subsequent returns in the same
direction. At negative $\rho$, these effects tend to offset. The
option prices lie above their frictionless references for positive
correlation and below them for negative correlation. Again the two
payoffs respond by similar percentages to the change in the
variability of the average.

In Figure \ref{fig:num-exog}(e)--(f), the additional value attributable
to order flow increases with $\lambda_I$ and decreases with $\sigma$
over the ranges plotted. Larger $\lambda_I$ increases the contribution
of order flow to returns. As $\sigma$ rises, ordinary price fluctuations
account for more of the uncertainty and the additional option value
from order flow falls in these calculations.

\subsection{Comparison of numerical methods}
\label{subsec:num-correctness}

We compare two approaches to the dealer's reservation prices. The
Bellman method works backwards from maturity, calculating continuation
values and selecting a trading rate at each discrete state. The neural
network methods approximate the value functions in the control equation.
Generalized policy iteration (GPI) alternates between evaluating a
trading rule and improving it, using physics-informed neural networks
\citep{Raissi2019,Dupret2026}. The deep Galerkin method (DGM)
fits a neural network directly to the control equation
\citep{Sirignano2018}.

The Policy column evaluates the GPI trading rules by Monte Carlo,
converting expected terminal utilities into reservation prices through
\eqref{eq:dealer-reservation-prices}. Table \ref{tab:num-correctness}
compares the methods for both payoffs, with and without dealer impact,
using the settings in Section \ref{subsec:num-setup}. The Bellman
columns report the baseline and extrapolated estimates.

\begin{table}[tbp]
\centering
\scriptsize
\setlength{\tabcolsep}{4pt}
\begin{adjustbox}{max width=\textwidth}
\begin{tabular}{@{}llrr|rrr@{}}
\hline
 & & \multicolumn{2}{c|}{\textbf{Bellman}}
   & \multicolumn{2}{c}{\textbf{Neural networks}} & \\
\cline{3-4}\cline{5-6}
$Q_0$ & & \textbf{Extrapolated} & \textbf{$N=25$}
  & \textbf{GPI} & \textbf{DGM} & \textbf{Policy} \\
\hline
\multicolumn{7}{@{}l}{\emph{Panel A: geometric Asian option, no dealer impact}} \\
0 & Bid    & $4.8760\pm0.0280$ & 4.9229 & \shortstack[r]{4.8798\\[-1pt]{\tiny $\pm$0.0542}} & \shortstack[r]{4.8459\\[-1pt]{\tiny $\pm$0.0470}} & \shortstack[r]{4.8689\\[-1pt]{\tiny $\pm$0.0038}} \\
1 & Ask    & $4.2552\pm0.0244$ & 4.4781 & \shortstack[r]{4.2366\\[-1pt]{\tiny $\pm$0.0320}} & \shortstack[r]{4.3004\\[-1pt]{\tiny $\pm$0.0498}} & \shortstack[r]{4.2636\\[-1pt]{\tiny $\pm$0.0036}} \\
1 & Bid    & $3.2340\pm0.0164$ & 3.4603 & \shortstack[r]{3.2480\\[-1pt]{\tiny $\pm$0.0293}} & \shortstack[r]{3.2306\\[-1pt]{\tiny $\pm$0.0525}} & \shortstack[r]{3.2306\\[-1pt]{\tiny $\pm$0.0030}} \\
1 & Spread & $1.0210\pm0.0145$ & 1.0178 & \shortstack[r]{0.9886\\[-1pt]{\tiny $\pm$0.0458}} & \shortstack[r]{1.0698\\[-1pt]{\tiny $\pm$0.0886}} & 1.0330 \\
2 & Ask    & $2.4374\pm0.0141$ & 2.7060 & \shortstack[r]{2.4215\\[-1pt]{\tiny $\pm$0.0203}} & \shortstack[r]{2.4798\\[-1pt]{\tiny $\pm$0.0211}} & \shortstack[r]{2.4406\\[-1pt]{\tiny $\pm$0.0033}} \\
2 & Bid    & $1.6814\pm0.0157$ & 1.8860 & \shortstack[r]{1.6869\\[-1pt]{\tiny $\pm$0.0150}} & \shortstack[r]{1.6715\\[-1pt]{\tiny $\pm$0.0493}} & \shortstack[r]{1.6812\\[-1pt]{\tiny $\pm$0.0025}} \\
2 & Spread & $0.7559\pm0.0066$ & 0.8200 & \shortstack[r]{0.7346\\[-1pt]{\tiny $\pm$0.0240}} & \shortstack[r]{0.8083\\[-1pt]{\tiny $\pm$0.0651}} & 0.7593 \\
\multicolumn{7}{@{}l}{\emph{Panel B: geometric Asian option, baseline dealer impact}} \\
0 & Bid    & $4.8749\pm0.0276$ & 4.9224 & \shortstack[r]{4.8740\\[-1pt]{\tiny $\pm$0.0620}} & \shortstack[r]{4.8551\\[-1pt]{\tiny $\pm$0.0471}} & \shortstack[r]{4.8678\\[-1pt]{\tiny $\pm$0.0037}} \\
1 & Ask    & $4.2512\pm0.0245$ & 4.4749 & \shortstack[r]{4.2409\\[-1pt]{\tiny $\pm$0.0462}} & \shortstack[r]{4.2307\\[-1pt]{\tiny $\pm$0.0404}} & \shortstack[r]{4.2593\\[-1pt]{\tiny $\pm$0.0036}} \\
1 & Bid    & $3.2294\pm0.0167$ & 3.4570 & \shortstack[r]{3.2417\\[-1pt]{\tiny $\pm$0.0482}} & \shortstack[r]{3.2396\\[-1pt]{\tiny $\pm$0.0438}} & \shortstack[r]{3.2269\\[-1pt]{\tiny $\pm$0.0030}} \\
1 & Spread & $1.0217\pm0.0150$ & 1.0179 & \shortstack[r]{0.9992\\[-1pt]{\tiny $\pm$0.0799}} & \shortstack[r]{0.9911\\[-1pt]{\tiny $\pm$0.0764}} & 1.0325 \\
2 & Ask    & $2.4322\pm0.0140$ & 2.7013 & \shortstack[r]{2.4197\\[-1pt]{\tiny $\pm$0.0545}} & \shortstack[r]{2.4332\\[-1pt]{\tiny $\pm$0.0247}} & \shortstack[r]{2.4354\\[-1pt]{\tiny $\pm$0.0033}} \\
2 & Bid    & $1.6779\pm0.0155$ & 1.8819 & \shortstack[r]{1.6757\\[-1pt]{\tiny $\pm$0.0495}} & \shortstack[r]{1.7023\\[-1pt]{\tiny $\pm$0.0265}} & \shortstack[r]{1.6787\\[-1pt]{\tiny $\pm$0.0025}} \\
2 & Spread & $0.7543\pm0.0064$ & 0.8195 & \shortstack[r]{0.7440\\[-1pt]{\tiny $\pm$0.0830}} & \shortstack[r]{0.7309\\[-1pt]{\tiny $\pm$0.0507}} & 0.7566 \\
\multicolumn{7}{@{}l}{\emph{Panel C: arithmetic Asian option, no dealer impact}} \\
0 & Bid    & $5.0579\pm0.0243$ & 5.1023 & \shortstack[r]{5.0207\\[-1pt]{\tiny $\pm$0.0656}} & \shortstack[r]{5.0184\\[-1pt]{\tiny $\pm$0.0601}} & \shortstack[r]{5.0481\\[-1pt]{\tiny $\pm$0.0039}} \\
1 & Ask    & $4.4549\pm0.0180$ & 4.6706 & \shortstack[r]{4.4418\\[-1pt]{\tiny $\pm$0.0902}} & \shortstack[r]{4.4822\\[-1pt]{\tiny $\pm$0.0382}} & \shortstack[r]{4.4592\\[-1pt]{\tiny $\pm$0.0037}} \\
1 & Bid    & $3.3688\pm0.0142$ & 3.5966 & \shortstack[r]{3.3363\\[-1pt]{\tiny $\pm$0.0834}} & \shortstack[r]{3.3458\\[-1pt]{\tiny $\pm$0.0706}} & \shortstack[r]{3.3619\\[-1pt]{\tiny $\pm$0.0031}} \\
1 & Spread & $1.0861\pm0.0090$ & 1.0740 & \shortstack[r]{1.1055\\[-1pt]{\tiny $\pm$0.1692}} & \shortstack[r]{1.1364\\[-1pt]{\tiny $\pm$0.0945}} & 1.0973 \\
2 & Ask    & $2.5715\pm0.0171$ & 2.8449 & \shortstack[r]{2.5571\\[-1pt]{\tiny $\pm$0.0598}} & \shortstack[r]{2.5830\\[-1pt]{\tiny $\pm$0.0534}} & \shortstack[r]{2.5720\\[-1pt]{\tiny $\pm$0.0034}} \\
2 & Bid    & $1.7606\pm0.0169$ & 1.9713 & \shortstack[r]{1.7286\\[-1pt]{\tiny $\pm$0.0795}} & \shortstack[r]{1.7436\\[-1pt]{\tiny $\pm$0.1046}} & \shortstack[r]{1.7588\\[-1pt]{\tiny $\pm$0.0026}} \\
2 & Spread & $0.8107\pm0.0122$ & 0.8736 & \shortstack[r]{0.8285\\[-1pt]{\tiny $\pm$0.1361}} & \shortstack[r]{0.8394\\[-1pt]{\tiny $\pm$0.1560}} & 0.8131 \\
\multicolumn{7}{@{}l}{\emph{Panel D: arithmetic Asian option, baseline dealer impact}} \\
0 & Bid    & $5.0568\pm0.0239$ & 5.1018 & \shortstack[r]{5.0654\\[-1pt]{\tiny $\pm$0.0657}} & \shortstack[r]{5.0033\\[-1pt]{\tiny $\pm$0.0371}} & \shortstack[r]{5.0467\\[-1pt]{\tiny $\pm$0.0039}} \\
1 & Ask    & $4.4509\pm0.0182$ & 4.6672 & \shortstack[r]{4.4254\\[-1pt]{\tiny $\pm$0.0426}} & \shortstack[r]{4.4791\\[-1pt]{\tiny $\pm$0.0523}} & \shortstack[r]{4.4555\\[-1pt]{\tiny $\pm$0.0037}} \\
1 & Bid    & $3.3639\pm0.0141$ & 3.5931 & \shortstack[r]{3.3713\\[-1pt]{\tiny $\pm$0.0250}} & \shortstack[r]{3.3313\\[-1pt]{\tiny $\pm$0.0438}} & \shortstack[r]{3.3568\\[-1pt]{\tiny $\pm$0.0031}} \\
1 & Spread & $1.0870\pm0.0094$ & 1.0742 & \shortstack[r]{1.0541\\[-1pt]{\tiny $\pm$0.0658}} & \shortstack[r]{1.1479\\[-1pt]{\tiny $\pm$0.0940}} & 1.0988 \\
2 & Ask    & $2.5659\pm0.0169$ & 2.8399 & \shortstack[r]{2.5295\\[-1pt]{\tiny $\pm$0.0666}} & \shortstack[r]{2.5813\\[-1pt]{\tiny $\pm$0.0660}} & \shortstack[r]{2.5666\\[-1pt]{\tiny $\pm$0.0034}} \\
2 & Bid    & $1.7569\pm0.0169$ & 1.9670 & \shortstack[r]{1.7685\\[-1pt]{\tiny $\pm$0.0327}} & \shortstack[r]{1.7387\\[-1pt]{\tiny $\pm$0.0688}} & \shortstack[r]{1.7549\\[-1pt]{\tiny $\pm$0.0026}} \\
2 & Spread & $0.8089\pm0.0119$ & 0.8729 & \shortstack[r]{0.7610\\[-1pt]{\tiny $\pm$0.0863}} & \shortstack[r]{0.8426\\[-1pt]{\tiny $\pm$0.1253}} & 0.8117 \\
\hline
\end{tabular}
\end{adjustbox}
\caption{Comparison of reservation bid and ask prices and their spread
for geometric (Panels A--B) and arithmetic (Panels C--D) Asian options.
All panels use $\lambda_I=0$. Panels A and C use $\bar\lambda_P=\bar\lambda_T=0$. Panels B and D use
$\bar\lambda_P=0.025$ and $\bar\lambda_T=0.05$. Bellman estimates are
reported at $N=25$ and after extrapolation, with estimated
extrapolation and discretization errors for the latter. GPI and DGM
entries are means over five training runs, with sample standard
deviations beneath them. Policy entries are Monte Carlo evaluations
of the trading rules from one GPI run, with standard errors beneath
the bid and ask.}
\label{tab:num-correctness}
\end{table}

With both dealer impact coefficients set to zero, increasing the initial
holding from one share to two lowers the extrapolated Bellman ask for
the geometric call from 4.26 to 2.44. Both neural network methods
produce nearly the same change. The bid also falls. Both quotes fall
with inventory for the arithmetic call and in the cases with dealer
impact. The spread estimates show larger relative differences across
methods than the individual quotes.

\subsection{Dealer quotes and trading}
\label{subsec:num-endogenous}
\label{subsec:num-three-model}

In this subsection we examine the cost of taking an option
position and the stock trades used to manage it. We use the
baseline $N=25$ calculation with $Q_0=J_0=0$.

\paragraph{Price comparisons.}
The dealer's price is the midpoint between the reservation bid and
ask, $P_{\rm dealer}=(P_{\rm ask}+P_{\rm bid})/2$.
Table \ref{tab:three-model} and Figure \ref{fig:num-three-model}
compare the exogenous price and the dealer's price relative to the
respective frictionless reference for each payoff. In Panels B and C, the
exogenous calculation uses $\lambda_I=0.05$, so $I_t$ affects the
expected return of $S_t$ and the distribution of the average
underlying the Asian payoff. For the dealer we use $\lambda_I=0$. Accordingly, a comparison of
these prices includes a difference in the market dynamics as well
as the effects of execution costs, constraints and risk preferences.
In Panel C the parameter varied is $\kappa_I$ in exogenous valuation
and $\kappa_J$ in the dealer calculation: external order flow decays
at rate $\kappa_I$, and the transient effect of the dealer's trades
decays at rate $\kappa_J$.

In the first row of Panel D, $\lambda_I=\bar\lambda_P=\bar\lambda_T=0$.
Each exogenous price agrees with its frictionless reference to the
precision reported. The dealer still faces temporary execution costs,
terminal liquidation costs and trading limits. Its finite-step price
can therefore differ from the reference.

\begin{table}[tbp]
\centering
\scriptsize
\setlength{\tabcolsep}{2.6pt}
\begin{adjustbox}{max width=\textwidth}
\begin{tabular}{@{}lrrrrrr|rrrrr@{}}
\hline
 & & \multicolumn{5}{c|}{\textbf{Geometric option}}
   & \multicolumn{5}{c}{\textbf{Arithmetic option}} \\
\cline{3-7}\cline{8-12}
\textbf{Para.} & \textbf{Value}
  & \begin{tabular}[b]{@{}c@{}}\textbf{KV}\\\textbf{Price}\end{tabular} & \begin{tabular}[b]{@{}c@{}}\textbf{Exog}\\\textbf{Price}\end{tabular} & \begin{tabular}[b]{@{}c@{}}\textbf{Exog}\\\textbf{\%}\end{tabular} & \begin{tabular}[b]{@{}c@{}}\textbf{Dealer}\\\textbf{Price}\end{tabular} & \begin{tabular}[b]{@{}c@{}}\textbf{Dealer}\\\textbf{\%}\end{tabular}
  & \begin{tabular}[b]{@{}c@{}}\textbf{Ref.}\\\textbf{Price}\end{tabular} & \begin{tabular}[b]{@{}c@{}}\textbf{Exog}\\\textbf{Price}\end{tabular} & \begin{tabular}[b]{@{}c@{}}\textbf{Exog}\\\textbf{\%}\end{tabular} & \begin{tabular}[b]{@{}c@{}}\textbf{Dealer}\\\textbf{Price}\end{tabular} & \begin{tabular}[b]{@{}c@{}}\textbf{Dealer}\\\textbf{\%}\end{tabular} \\
\hline
\multicolumn{7}{@{}l|}{\emph{Panel A: dealer transient impact $\bar\lambda_T$ ($\bar\lambda_P=\lambda_I=0$)}} & \multicolumn{5}{l}{} \\
$\bar\lambda_T$ & 0.00 & 5.5468 & 5.5468 & 0.000 & 5.6772 & 2.351 & \shortstack[r]{5.7647\\[-1pt]{\tiny $\pm$0.0015}} & \shortstack[r]{5.7647\\[-1pt]{\tiny $\pm$0.0015}} & 0.000 & 5.8930 & 2.226 \\
$\bar\lambda_T$ & 0.02 & 5.5468 & 5.5468 & 0.000 & 5.6773 & 2.353 & \shortstack[r]{5.7647\\[-1pt]{\tiny $\pm$0.0015}} & \shortstack[r]{5.7647\\[-1pt]{\tiny $\pm$0.0015}} & 0.000 & 5.8931 & 2.227 \\
$\bar\lambda_T$ & 0.05 & 5.5468 & 5.5468 & 0.000 & 5.6774 & 2.354 & \shortstack[r]{5.7647\\[-1pt]{\tiny $\pm$0.0015}} & \shortstack[r]{5.7647\\[-1pt]{\tiny $\pm$0.0015}} & 0.000 & 5.8932 & 2.229 \\
$\bar\lambda_T$ & 0.10 & 5.5468 & 5.5468 & 0.000 & 5.6775 & 2.356 & \shortstack[r]{5.7647\\[-1pt]{\tiny $\pm$0.0015}} & \shortstack[r]{5.7647\\[-1pt]{\tiny $\pm$0.0015}} & 0.000 & 5.8933 & 2.232 \\
$\bar\lambda_T$ & 0.15 & 5.5468 & 5.5468 & 0.000 & 5.6776 & 2.358 & \shortstack[r]{5.7647\\[-1pt]{\tiny $\pm$0.0015}} & \shortstack[r]{5.7647\\[-1pt]{\tiny $\pm$0.0015}} & 0.000 & 5.8935 & 2.234 \\
$\bar\lambda_T$ & 0.20 & 5.5468 & 5.5468 & 0.000 & 5.6778 & 2.360 & \shortstack[r]{5.7647\\[-1pt]{\tiny $\pm$0.0015}} & \shortstack[r]{5.7647\\[-1pt]{\tiny $\pm$0.0015}} & 0.000 & 5.8936 & 2.237 \\
\multicolumn{7}{@{}l|}{\emph{Panel B: volatility $\sigma$}} & \multicolumn{5}{l}{} \\
$\sigma$ & 0.10 & 3.5723 & 3.5786 & 0.176 & 3.6165 & 1.240 & \shortstack[r]{3.6419\\[-1pt]{\tiny $\pm$0.0004}} & \shortstack[r]{3.6486\\[-1pt]{\tiny $\pm$0.0004}} & 0.184 & 3.6864 & 1.222 \\
$\sigma$ & 0.15 & 4.5542 & 4.5588 & 0.100 & 4.6416 & 1.917 & \shortstack[r]{4.6873\\[-1pt]{\tiny $\pm$0.0009}} & \shortstack[r]{4.6922\\[-1pt]{\tiny $\pm$0.0009}} & 0.104 & 4.7742 & 1.854 \\
$\sigma$ & 0.20 & 5.5468 & 5.5504 & 0.065 & 5.6774 & 2.355 & \shortstack[r]{5.7647\\[-1pt]{\tiny $\pm$0.0015}} & \shortstack[r]{5.7686\\[-1pt]{\tiny $\pm$0.0016}} & 0.068 & 5.8932 & 2.230 \\
$\sigma$ & 0.30 & 7.4960 & 7.4985 & 0.034 & 7.7166 & 2.944 & \shortstack[r]{7.9487\\[-1pt]{\tiny $\pm$0.0035}} & \shortstack[r]{7.9516\\[-1pt]{\tiny $\pm$0.0035}} & 0.037 & 8.1574 & 2.626 \\
$\sigma$ & 0.40 & 9.3650 & 9.3671 & 0.022 & 9.6629 & 3.181 & \shortstack[r]{10.1405\\[-1pt]{\tiny $\pm$0.0065}} & \shortstack[r]{10.1429\\[-1pt]{\tiny $\pm$0.0065}} & 0.024 & 10.4073 & 2.631 \\
\multicolumn{7}{@{}l|}{\emph{Panel C: mean reversion $\kappa$ ($\kappa_I$ exogenous, $\kappa_J$ endogenous)}} & \multicolumn{5}{l}{} \\
$\kappa$ & 0.50 & 5.5468 & 5.5514 & 0.083 & 5.6774 & 2.354 & \shortstack[r]{5.7647\\[-1pt]{\tiny $\pm$0.0015}} & \shortstack[r]{5.7697\\[-1pt]{\tiny $\pm$0.0016}} & 0.088 & 5.8932 & 2.229 \\
$\kappa$ & 1.00 & 5.5468 & 5.5504 & 0.065 & 5.6774 & 2.355 & \shortstack[r]{5.7647\\[-1pt]{\tiny $\pm$0.0015}} & \shortstack[r]{5.7686\\[-1pt]{\tiny $\pm$0.0016}} & 0.068 & 5.8932 & 2.230 \\
$\kappa$ & 2.00 & 5.5468 & 5.5491 & 0.042 & 5.6775 & 2.356 & \shortstack[r]{5.7647\\[-1pt]{\tiny $\pm$0.0015}} & \shortstack[r]{5.7672\\[-1pt]{\tiny $\pm$0.0015}} & 0.044 & 5.8933 & 2.231 \\
$\kappa$ & 5.00 & 5.5468 & 5.5477 & 0.015 & 5.6775 & 2.356 & \shortstack[r]{5.7647\\[-1pt]{\tiny $\pm$0.0015}} & \shortstack[r]{5.7656\\[-1pt]{\tiny $\pm$0.0015}} & 0.016 & 5.8933 & 2.231 \\
$\kappa$ & 10.00 & 5.5468 & 5.5471 & 0.005 & 5.6775 & 2.356 & \shortstack[r]{5.7647\\[-1pt]{\tiny $\pm$0.0015}} & \shortstack[r]{5.7650\\[-1pt]{\tiny $\pm$0.0015}} & 0.005 & 5.8932 & 2.230 \\
\multicolumn{7}{@{}l|}{\emph{Panel D: dealer permanent impact $\bar\lambda_P$ ($\bar\lambda_T=\lambda_I=0$)}} & \multicolumn{5}{l}{} \\
$\bar\lambda_P$ & 0.000 & 5.5468 & 5.5468 & 0.000 & 5.6772 & 2.351 & \shortstack[r]{5.7647\\[-1pt]{\tiny $\pm$0.0015}} & \shortstack[r]{5.7647\\[-1pt]{\tiny $\pm$0.0015}} & 0.000 & 5.8930 & 2.226 \\
$\bar\lambda_P$ & 0.025 & 5.5468 & 5.5468 & 0.000 & 5.6773 & 2.352 & \shortstack[r]{5.7647\\[-1pt]{\tiny $\pm$0.0015}} & \shortstack[r]{5.7647\\[-1pt]{\tiny $\pm$0.0015}} & 0.000 & 5.8930 & 2.227 \\
$\bar\lambda_P$ & 0.050 & 5.5468 & 5.5468 & 0.000 & 5.6773 & 2.353 & \shortstack[r]{5.7647\\[-1pt]{\tiny $\pm$0.0015}} & \shortstack[r]{5.7647\\[-1pt]{\tiny $\pm$0.0015}} & 0.000 & 5.8931 & 2.228 \\
$\bar\lambda_P$ & 0.100 & 5.5468 & 5.5468 & 0.000 & 5.6774 & 2.354 & \shortstack[r]{5.7647\\[-1pt]{\tiny $\pm$0.0015}} & \shortstack[r]{5.7647\\[-1pt]{\tiny $\pm$0.0015}} & 0.000 & 5.8932 & 2.230 \\
$\bar\lambda_P$ & 0.200 & 5.5468 & 5.5468 & 0.000 & 5.6775 & 2.356 & \shortstack[r]{5.7647\\[-1pt]{\tiny $\pm$0.0015}} & \shortstack[r]{5.7647\\[-1pt]{\tiny $\pm$0.0015}} & 0.000 & 5.8934 & 2.232 \\
\multicolumn{7}{@{}l|}{\emph{Panel E: drift loading $\lambda_I$ ($\bar\lambda_T=\bar\lambda_P=0$)}} & \multicolumn{5}{l}{} \\
$\lambda_I$ & 0.00 & 5.5468 & 5.5468 & 0.000 & 5.6772 & 2.351 & \shortstack[r]{5.7647\\[-1pt]{\tiny $\pm$0.0015}} & \shortstack[r]{5.7647\\[-1pt]{\tiny $\pm$0.0015}} & 0.000 & 5.8930 & 2.226 \\
$\lambda_I$ & 0.05 & 5.5468 & 5.5504 & 0.065 & 5.7045 & 2.842 & \shortstack[r]{5.7647\\[-1pt]{\tiny $\pm$0.0015}} & \shortstack[r]{5.7686\\[-1pt]{\tiny $\pm$0.0016}} & 0.068 & 5.9279 & 2.831 \\
$\lambda_I$ & 0.10 & 5.5468 & 5.5612 & 0.258 & 5.7322 & 3.343 & \shortstack[r]{5.7647\\[-1pt]{\tiny $\pm$0.0015}} & \shortstack[r]{5.7805\\[-1pt]{\tiny $\pm$0.0016}} & 0.275 & 5.9606 & 3.398 \\
$\lambda_I$ & 0.15 & 5.5468 & 5.5790 & 0.581 & 5.7599 & 3.841 & \shortstack[r]{5.7647\\[-1pt]{\tiny $\pm$0.0015}} & \shortstack[r]{5.8003\\[-1pt]{\tiny $\pm$0.0016}} & 0.618 & 5.9977 & 4.042 \\
$\lambda_I$ & 0.20 & 5.5468 & 5.6040 & 1.030 & 5.7880 & 4.347 & \shortstack[r]{5.7647\\[-1pt]{\tiny $\pm$0.0015}} & \shortstack[r]{5.8279\\[-1pt]{\tiny $\pm$0.0016}} & 1.098 & 6.0368 & 4.721 \\
\hline
\end{tabular}
\end{adjustbox}
\caption{Exogenous prices and the dealer's prices.
KV Price is the geometric Kemna--Vorst price, and Ref.\ Price is
the arithmetic Monte Carlo reference. Dealer Price is $P_{\rm dealer}$.
Exog\ \% and Dealer\ \% are percentage
differences from the respective frictionless reference. Arithmetic Monte Carlo: nominal $95\%$ sampling half widths,
without averaging or control-variate centring errors. Finite step Bellman prices use $N=25$, $Q_0=J_0=0$ and $n_I=9$,
with $\lambda_I=0$ except in Panel E. Panel C varies $\kappa_I$
in the exogenous calculation and $\kappa_J$ in the dealer calculation.}
\label{tab:three-model}
\end{table}

\begin{figure}[tbp]
\centering
\includegraphics[width=\textwidth]{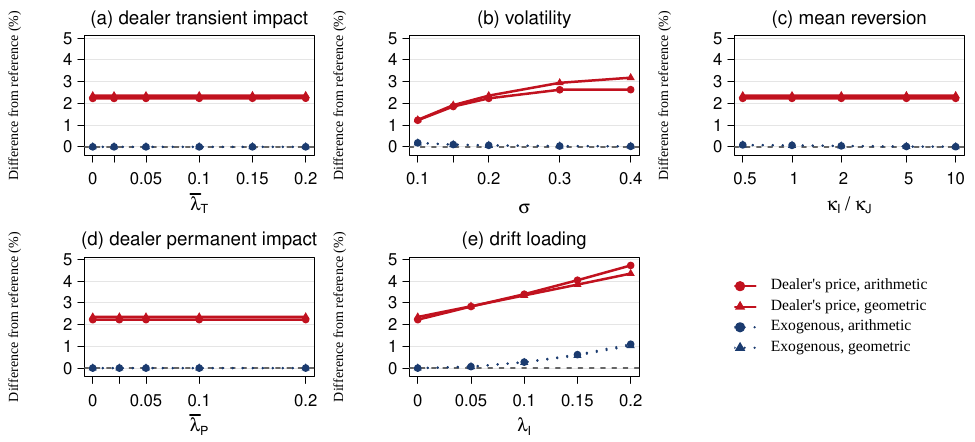}
\caption{Percentage differences of exogenous prices (blue) and the dealer's
prices (red) from their frictionless references,
using Table \ref{tab:three-model}. Circles denote arithmetic options
and triangles geometric options. Dealer calculations use $N=25$
and $Q_0=0$.}
\label{fig:num-three-model}
\end{figure}

Panels A, C and D show little sensitivity of the dealer's price to
$\bar\lambda_T$, $\kappa_J$ and $\bar\lambda_P$, respectively.
These parameters affect the hedge through its execution payments.
The distribution of the payoff is unchanged. At the reported
resolution, this effect is small relative to the variations with
volatility and external order flow over the ranges in the table.

In Panel B, both option prices increase with $\sigma$, as expected
from the effect of uncertainty on a call payoff. Their percentage
premiums over the frictionless reference behave differently. The
exogenous premium is small and falls: ordinary price fluctuations
become larger relative to the contribution of order flow. The
dealer's premium is larger and increases. With costly and limited
trading, the hedge cannot be adjusted freely as volatility rises.

Panel E varies $\lambda_I$ with both dealer impact coefficients set to zero.
Increasing $\lambda_I$ raises both prices, with a
larger increase in the dealer's price. External order flow affects
the distribution of the payoff and the dealer's subsequent hedge
adjustments.

\paragraph{Prices and spreads.}
\label{subsec:num-spread}
Table \ref{tab:spread-sens} and Figure \ref{fig:num-indiff} show how
hedging costs and risk preferences affect the dealer's prices and spreads.
The dealer's prices and absolute spreads are higher for arithmetic options
throughout the table. Their larger payoffs are accompanied by a larger
difference between the dealer's selling and buying valuations.
Relative to the dealer's price, however, the spreads for the two payoffs
are very similar. At the baseline parameters, both are slightly
more than a quarter of the dealer's price.

\begin{table}[tbp]
\centering
\scriptsize
\setlength{\tabcolsep}{4pt}
\begin{tabular}{@{}llrrrr|rrrr@{}}
\hline
 & & \multicolumn{4}{c|}{\textbf{Geometric option}} & \multicolumn{4}{c}{\textbf{Arithmetic option}} \\
\cline{3-6}\cline{7-10}
\textbf{Par.} & \textbf{Value} & \textbf{Price} & \textbf{Spread} & \textbf{Sp.\ \%} & \textbf{Imp.\ \%}
  & \textbf{Price} & \textbf{Spread} & \textbf{Sp.\ \%} & \textbf{Imp.\ \%} \\
\hline
\multicolumn{6}{@{}l|}{\emph{Panel A: cost level $k=k_A=k_B$}} & \multicolumn{4}{l}{} \\
$k$ & 0.001 & 5.6347 & 0.1550 & 2.75 & 1.585 & 5.8469 & 0.1644 & 2.81 & 1.427 \\
$k$ & 0.010 & 5.6347 & 0.2359 & 4.19 & 1.584 & 5.8468 & 0.2481 & 4.24 & 1.425 \\
$k$ & 0.100 & 5.6402 & 0.7138 & 12.65 & 1.683 & 5.8522 & 0.7435 & 12.70 & 1.518 \\
$k$ & 0.500 & 5.6774 & 1.5100 & 26.60 & 2.355 & 5.8932 & 1.5828 & 26.86 & 2.230 \\
$k$ & 1.000 & 5.7132 & 1.9175 & 33.56 & 3.000 & 5.9322 & 2.0171 & 34.00 & 2.907 \\
\multicolumn{6}{@{}l|}{\emph{Panel B: $\bar\lambda_T$ ($\bar\lambda_P=\bar\lambda_T/2$)}} & \multicolumn{4}{l}{} \\
$\bar\lambda_T$ & 0.00 & 5.6772 & 1.5086 & 26.57 & 2.351 & 5.8930 & 1.5814 & 26.83 & 2.226 \\
$\bar\lambda_T$ & 0.05 & 5.6774 & 1.5100 & 26.60 & 2.355 & 5.8932 & 1.5828 & 26.86 & 2.230 \\
$\bar\lambda_T$ & 0.10 & 5.6776 & 1.5111 & 26.62 & 2.357 & 5.8934 & 1.5840 & 26.88 & 2.233 \\
$\bar\lambda_T$ & 0.20 & 5.6779 & 1.5129 & 26.65 & 2.363 & 5.8938 & 1.5860 & 26.91 & 2.240 \\
$\bar\lambda_T$ & 0.30 & 5.6783 & 1.5140 & 26.66 & 2.370 & 5.8942 & 1.5872 & 26.93 & 2.247 \\
\multicolumn{6}{@{}l|}{\emph{Panel C: $\bar\lambda_P$ ($\bar\lambda_T=0.05$)}} & \multicolumn{4}{l}{} \\
$\bar\lambda_P$ & 0.000 & 5.6774 & 1.5106 & 26.61 & 2.354 & 5.8932 & 1.5835 & 26.87 & 2.229 \\
$\bar\lambda_P$ & 0.025 & 5.6774 & 1.5100 & 26.60 & 2.355 & 5.8932 & 1.5828 & 26.86 & 2.230 \\
$\bar\lambda_P$ & 0.050 & 5.6775 & 1.5093 & 26.58 & 2.355 & 5.8933 & 1.5821 & 26.85 & 2.231 \\
$\bar\lambda_P$ & 0.100 & 5.6775 & 1.5080 & 26.56 & 2.356 & 5.8934 & 1.5807 & 26.82 & 2.232 \\
$\bar\lambda_P$ & 0.200 & 5.6776 & 1.5052 & 26.51 & 2.359 & 5.8935 & 1.5778 & 26.77 & 2.235 \\
\multicolumn{6}{@{}l|}{\emph{Panel D: decay $\kappa_J$}} & \multicolumn{4}{l}{} \\
$\kappa_J$ & 0.50 & 5.6774 & 1.5093 & 26.58 & 2.354 & 5.8932 & 1.5820 & 26.84 & 2.229 \\
$\kappa_J$ & 1.00 & 5.6774 & 1.5100 & 26.60 & 2.355 & 5.8932 & 1.5828 & 26.86 & 2.230 \\
$\kappa_J$ & 2.00 & 5.6775 & 1.5108 & 26.61 & 2.356 & 5.8933 & 1.5838 & 26.87 & 2.231 \\
$\kappa_J$ & 5.00 & 5.6775 & 1.5114 & 26.62 & 2.356 & 5.8933 & 1.5843 & 26.88 & 2.231 \\
$\kappa_J$ & 10.00 & 5.6775 & 1.5108 & 26.61 & 2.356 & 5.8932 & 1.5837 & 26.87 & 2.230 \\
\multicolumn{6}{@{}l|}{\emph{Panel E: cost exponent $\psi$}} & \multicolumn{4}{l}{} \\
$\psi$ & 0.20 & 5.6413 & 1.1655 & 20.66 & 1.703 & 5.8536 & 1.2096 & 20.66 & 1.543 \\
$\psi$ & 0.40 & 5.6506 & 1.3011 & 23.03 & 1.870 & 5.8625 & 1.3533 & 23.08 & 1.697 \\
$\psi$ & 0.60 & 5.6618 & 1.3945 & 24.63 & 2.073 & 5.8758 & 1.4565 & 24.79 & 1.928 \\
$\psi$ & 0.80 & 5.6732 & 1.4607 & 25.75 & 2.278 & 5.8872 & 1.5280 & 25.95 & 2.126 \\
$\psi$ & 1.00 & 5.6774 & 1.5100 & 26.60 & 2.355 & 5.8932 & 1.5828 & 26.86 & 2.230 \\
\multicolumn{6}{@{}l|}{\emph{Panel F: risk aversion $\gamma$}} & \multicolumn{4}{l}{} \\
$\gamma$ & 0.010 & 5.6286 & 0.4889 & 8.69 & 1.475 & 5.8408 & 0.5169 & 8.85 & 1.320 \\
$\gamma$ & 0.020 & 5.6387 & 0.8274 & 14.67 & 1.657 & 5.8512 & 0.8715 & 14.89 & 1.500 \\
$\gamma$ & 0.050 & 5.6774 & 1.5100 & 26.60 & 2.355 & 5.8932 & 1.5828 & 26.86 & 2.230 \\
$\gamma$ & 0.075 & 5.7134 & 1.9140 & 33.50 & 3.002 & 5.9289 & 2.0037 & 33.80 & 2.848 \\
$\gamma$ & 0.100 & 5.7430 & 2.2532 & 39.23 & 3.537 & 5.9618 & 2.3583 & 39.56 & 3.419 \\
\hline
\end{tabular}
\caption{Dealer's prices and spreads for geometric and arithmetic Asian options
from the finite step Bellman calculation, with $N=25$, $Q_0=J_0=0$
and $\lambda_I=0$.
Price is $P_{\rm dealer}$ and Spread is
$P_{\rm ask}-P_{\rm bid}$. Sp.\ \% is
$100\mathrm{Spread}/P_{\rm dealer}$, and Imp.\ \% is
$100(P_{\rm dealer}-P_{\rm ref})/P_{\rm ref}$. The reference $P_{\rm ref}$
is the Kemna--Vorst price $5.5468$ for the geometric option and the
Monte Carlo estimate $5.7647$ for the arithmetic option.
Parameters follow Section \ref{subsec:num-setup} except where indicated.}
\label{tab:spread-sens}
\end{table}

\begin{figure}[tbp]
\centering
\includegraphics[width=\textwidth]{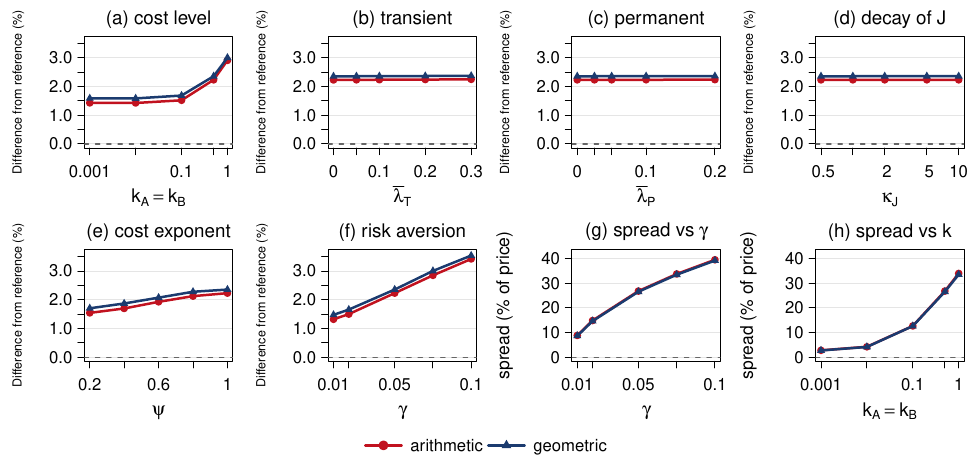}
\caption{Arithmetic and geometric Asian option quote sensitivities from
Table \ref{tab:spread-sens}. (a)--(f) Percentage differences of the
dealer's price from the respective frictionless reference. (g)--(h) Spread
as a percentage of the dealer's price, varying risk aversion $\gamma$ and
temporary cost $k=k_A=k_B$. }
\label{fig:num-indiff}
\end{figure}

Temporary costs and risk aversion produce the largest changes in
spreads over the ranges studied. In Panel A, higher $k=k_A=k_B$
substantially widens the spread for both payoffs. More expensive
stock adjustments increase the cost of hedging the option. A
positive spread remains even at the lowest cost level. The dealer's price changes much less than the spread.

Higher risk aversion in Panel F also widens the spread substantially,
with the ask rising and the bid falling. The dealer requires greater
compensation to sell the option and is willing to pay less to buy it,
as the hedge leaves residual risk. This effect is much larger than
the accompanying increase in the dealer's price, and is similar for the
two payoff conventions.

Increasing the cost exponent $\psi$ in Panel E raises both the
dealer's price and the spread, with a smaller effect on the spread than
the variations in cost level or risk aversion. Under $C(\nu)$, a
larger exponent increases the cost of rates with $|\nu|>1$, making
rapid hedge adjustments more expensive. At the reported resolution,
changes in the dealer's impact coefficients produce much smaller
changes in the calculated quotes than changes in temporary costs
or risk aversion. Changes in the decay rate in Panel D also have
little effect.

\paragraph{Trading paths.}
\label{subsec:num-hedge}
Table \ref{tab:hedge} follows the hedge with both price and order
flow innovations equal to zero. The option seller initially buys
stock, and the option buyer sells it. For both payoffs the stock
position is largely built at the beginning and reduced as fewer
future prices remain in the average. Costs make repeated large
adjustments expensive, consistent with the gradual changes along
these paths. The remaining inventory at expiry is subject to the
terminal liquidation rule of Section \ref{sec:endogenous}.

\begin{table}[tbp]
\centering
\scriptsize
\setlength{\tabcolsep}{4pt}
\begin{tabular}{rrrrrrrrrr}
\hline
 & & \multicolumn{4}{c}{\textbf{Trading rate $\nu_{m}$}}
   & \multicolumn{4}{c}{\textbf{Inventory $Q_{t_m}$}} \\
\cline{3-6}\cline{7-10}
$m$ & $t_m$
  & \textbf{Ar.\ sell} & \textbf{Ar.\ buy} & \textbf{Ge.\ sell} & \textbf{Ge.\ buy}
  & \textbf{Ar.\ sell} & \textbf{Ar.\ buy} & \textbf{Ge.\ sell} & \textbf{Ge.\ buy} \\
\hline
0 & 0.00 & 2.1429 & $-$2.1429 & 2.1429 & $-$1.4286 & 0.0000 & 0.0000 & 0.0000 & 0.0000 \\
5 & 0.20 & 0.3984 & $-$0.4093 & 0.3980 & $-$0.4834 & 0.2442 & $-$0.2427 & 0.2443 & $-$0.2323 \\
10 & 0.40 & 0.0000 & 0.0000 & 0.0000 & 0.0000 & 0.2790 & $-$0.2701 & 0.2734 & $-$0.2540 \\
15 & 0.60 & $-$0.3467 & 0.3075 & $-$0.2524 & 0.1856 & 0.2485 & $-$0.2430 & 0.2353 & $-$0.2260 \\
20 & 0.80 & $-$0.4413 & 0.5023 & $-$0.0995 & 0.5105 & 0.1943 & $-$0.1703 & 0.1988 & $-$0.1715 \\
23 & 0.92 & 0.0000 & 0.1830 & 0.0000 & 0.1861 & 0.1743 & $-$0.1256 & 0.1924 & $-$0.1260 \\
24 & 0.96 & 0.0000 & 0.1307 & 0.0000 & 0.1329 & 0.1743 & $-$0.1183 & 0.1924 & $-$0.1186 \\
25 & 1.00 & --- & --- & --- & --- & 0.1743 & $-$0.1131 & 0.1924 & $-$0.1133 \\
\hline
\end{tabular}
\caption{Trading rates and inventories for the option seller
($\theta=+1$) and buyer ($\theta=-1$), with $K=100$ and $Q_0=J_0=0$.
Price and order flow innovations are zero along this deterministic
path. A positive rate is a purchase. Ar.\ denotes the arithmetic
option and Ge.\ the geometric option. Trading rates between stored
inventory values are obtained by interpolation.}
\label{tab:hedge}
\end{table}

\begin{figure}[tbp]
\centering
\includegraphics[width=\textwidth]{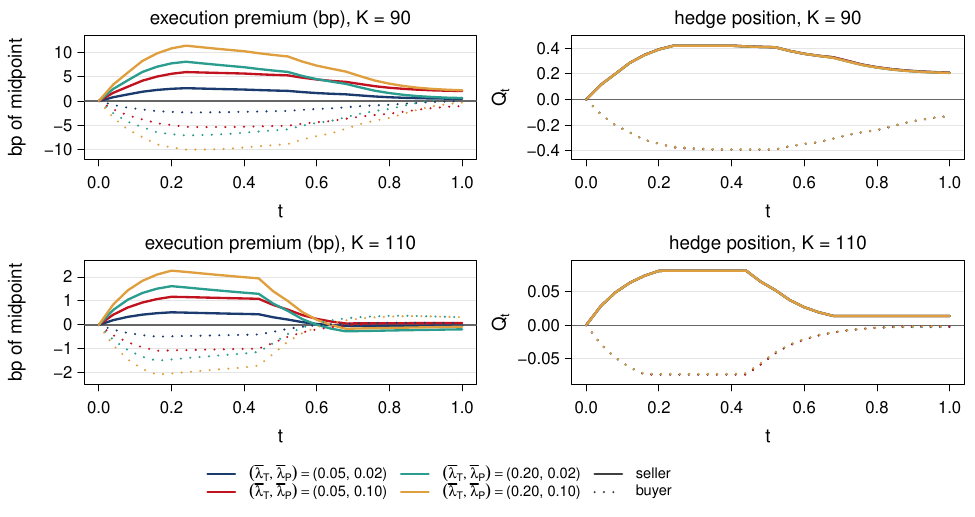}
\caption{Execution adjustment and inventory for geometric Asian options at
$K=90$ (top) and $K=110$ (bottom), along the deterministic paths
described in the text. The left panels show
$10^4(\bar\lambda_PQ_t+\bar\lambda_TJ_t)/S_t$ in basis points.}
\label{fig:num-trajectories}
\end{figure}

The paths at $K=110$ in Figure \ref{fig:num-trajectories} have smaller
stock holdings than those at $K=90$. Along these paths, the payoff at
the higher strike is less sensitive to subsequent price changes.
The difference between the execution and market prices,
$\widehat S_t-S_t=\bar\lambda_PQ_t+\bar\lambda_TJ_t$, is also plotted.
It includes the inventory adjustment and the transient effect of past
trades. Increasing the impact coefficients enlarges this difference
but has only a small effect on the plotted stock holdings. Cash paid
for the hedge uses $\widehat S_t$, whereas the payoff uses $S_t$.

\section{Discussion}
\label{sec:discussion}

We have studied arithmetic and geometric Asian options with external
order flow in the market dynamics and execution costs in the dealer's
hedge. Exogenous valuation and utility-indifference valuation use the
same payoff price, which is independent of the dealer's trades. The
execution price depends on its inventory and trading history.

For the market model, the discrete processes converge to a diffusion.
The spread condition permits an explicit consistent price system. Under the resulting measure, the arithmetic
option satisfies a pricing PDE and the geometric option admits an
explicit formula. Initial order flow, mean reversion and correlation appear explicitly
in the geometric Asian option formula.

For the dealer, utility indifference compares terminal cash and stock
wealth after execution payments and liquidation. From this comparison
we obtain the bid and ask, derive the formal control equation and
compute finite step values by Bellman iteration. Arbitrage in stock
trading is excluded for zero initial endowment. In the frictionless case specified above, both reservation prices
equal the replication price.

The numerical experiments show lower required payments for selling
a call when the dealer starts with stock. The liability offsets part
of the stock exposure. In the finite-step calculations from zero
inventory, higher temporary costs or risk aversion widen the spread. These quotes can be compared with the exogenous price for different
initial inventories, execution costs and trading limits.

Other averaging dates and time-dependent execution costs and trading
capacity are possible extensions. A portfolio of Asian options with
different strikes and maturities would use a common stock hedge.
Reservation prices would then depend on offsetting option exposures
and on the allocation of limited trading capacity across contracts.

\subsection*{Disclosure statement}

The opinions expressed in this work are solely those of the authors
and do not represent those of their current or past employers.
The authors report no potential conflict of interest.

\subsection*{Funding}

No funding was received for this work.

\subsection*{Statement on AI usage}

GPT-6 Astra was used in the proofreading of this paper, including the verification of mathematical results. Claude Opus 5.5 was used to generate the code for the numerical implementations in the paper.

\printbibliography

\appendix
\section{Finiteness of the utility problem}
\label{subsec:dealer-wellposedness}

The no-arbitrage proof in Section \ref{subsec:dealer-price-comparisons}
uses the following discounted impact costs. From integration by parts
with $d(u)=e^{-r(u-t)}$ and \eqref{eq:dealer-states}, we obtain
\begin{align}
    \int_t^T d(u)\bar\lambda_PQ_u\nu_u du
    &=\frac{\bar\lambda_P}{2}
       \left[d(T)Q_T^2-q^2+r\int_t^T d(u)Q_u^2du\right],
    \label{eq:dealer-permanent-cost}\\
    \int_t^T d(u)\bar\lambda_TJ_u\nu_u du
    &=\frac{\bar\lambda_T}{2}\left[d(T)J_T^2-j^2\right]
       +\bar\lambda_T\left(\kappa_J+\frac r2\right)
          \int_t^T d(u)J_u^2du.
    \label{eq:dealer-transient-cost}
\end{align}
Starting from $Q_t=J_t=0$, the total discounted impact charge is
nonnegative when $r\geq0$, even if an individual trade earns a rebate.

A sufficient condition on the initial inventory for finite reservation
prices is
\[
q\geq q_{\mathrm{cov}}(t)
   =\frac nT\int_t^T e^{-r(T-u)}du,
\]
under the trading bounds in the following proposition.
\begin{prop}[Finiteness with covering inventory]
\label{prop:dealer-covered}
Suppose
\[
\bar\nu\geq n/T,\qquad q_{\mathrm{cov}}(t)\leq q\leq\bar Q.
\]
For either Asian call option and any finite initial state
$(x,s,i,q,j,a)$ with $s>0$ and $a=0$ when $t=0$, all three utility
values lie in $(-\infty,0)$. The continuous time reservation bid
and ask are therefore finite.
\end{prop}

\begin{proof}
The rate bound implies $|J_u|\leq |j|+\bar\nu/\kappa_J$.
The execution adjustment is bounded in absolute value by
\[
M=\bar\lambda_P\bar Q
  +\bar\lambda_T\left(|j|+\frac{\bar\nu}{\kappa_J}\right).
\]
The sale rate
\[
\nu_u=-\frac nT e^{-r(T-u)}\mathbf 1_{\{S_u>M+\varepsilon\}}
\]
satisfies $|\nu_u|\leq n/T\leq\bar\nu$, since $r\geq0$. Its inventory
satisfies
\[
0\leq q-q_{\mathrm{cov}}(t)\leq Q_u\leq q\leq\bar Q,
\qquad t\leq u\leq T.
\]
At each sale, $\widehat S_u\geq S_u-M>\varepsilon$, so the execution
constraint also holds.
The midpoint proceeds carried to maturity satisfy
\[
 -\int_t^T e^{r(T-u)}S_u\nu_u du
 \geq \frac nT\int_t^T S_u du-\frac{n(T-t)}T(M+\varepsilon).
\]
For $t>0$, the payoff bounds are
\[
\begin{aligned}
 \Phi_A(R_T)&\leq\frac1T\int_t^T S_u du+\left(\frac aT-K\right)^+,\\
 \Phi_G(R_T)&\leq\frac1T\int_t^T S_u du+\frac tT e^{a/t}.
\end{aligned}
\]
The geometric bound follows from the weighted arithmetic--geometric
mean inequality. At $t=0$, both payoffs are bounded by the arithmetic
average. The remaining trading costs are bounded and $Q_TS_T\geq0$,
so $W_T^{+1,\nu}\geq-c$ for a finite deterministic constant $c$
depending on the initial state.
The same strategy satisfies $W_T^{0,\nu},W_T^{-1,\nu}\geq W_T^{+1,\nu}$.
Hence all three utility values exceed $-\infty$.

The Gaussian log-price dynamics imply
$\mathbb E^{\mathbb P}[\sup_{t\leq u\leq T}S_u]<\infty$.
The rate and inventory bounds therefore imply
$\sup_\nu\mathbb E^{\mathbb P}[W_T^{\theta,\nu}]\leq c_1<\infty$,
with $c_1$ independent of the admissible strategy. By Jensen's
inequality,
\[
-e^{\gamma c}\leq u^\theta\leq-e^{-\gamma c_1}<0,
\qquad \theta\in\{-1,0,+1\}.
\]
\end{proof}

The baseline calculations in Section \ref{sec:numerics} use $Q_0=0$.
The seller's expected exponential loss may be infinite for a call
option on a lognormal asset when its risk cannot be fully hedged.
\citet[p. 369]{henderson2002claims} notes this for a claim on a
nontraded asset. In a bounded-rate CARA hedging model,
\citet[Remark 2.2]{GueantPu2017} use Bachelier stock dynamics to avoid
this concern. For the present Asian call options, this integrability problem occurs
under the zero inventory conditions below.
\begin{prop}[Zero initial inventory and exponential utility]
\label{prop:dealer-zero-inventory}
Suppose $\lambda_I=0$, $r\geq0$, $S_0=s>0$, and
$Q_0=J_0=R_0=0$. For either uncapped Asian call option, every admissible
strategy and every finite initial cash amount satisfy
\[
    \mathbb E^{\mathbb P}[e^{-\gamma W_T^{+1,\nu}}]=\infty,
    \qquad \gamma>0.
\]
Thus $u^{+1}=-\infty$, and no finite payment satisfies the
reservation-ask equality at this initial state.
\end{prop}

\begin{proof}
For $\bar S_u=e^{-ru}S_u$, integration by parts and $Q_0=0$ imply
\[
\int_0^T Q_u d\bar S_u
 =Q_T\bar S_T-\int_0^T\bar S_u\nu_u du
 =\int_0^T\nu_u(\bar S_T-\bar S_u)du.
\]
By \eqref{eq:dealer-permanent-cost} and
\eqref{eq:dealer-transient-cost}, total discounted impact costs are
nonnegative when $Q_0=J_0=0$ and $r\geq0$. Temporary and terminal
costs are also nonnegative. The rate bound therefore implies
\begin{equation}
 W_T^{+1,\nu}
 \leq e^{rT}x+e^{rT}\bar\nu\int_0^T|\bar S_T-\bar S_u|du
          -n\Phi(R_T).
 \label{eq:zero-inventory-wealth-bound}
\end{equation}

Since $\lambda_I=0$, the discounted midpoint satisfies
$\log(\bar S_u/s)=b(u)+\sigma W_u$, where
$b(u)=\int_0^u(\mu(v)-r-\sigma^2/2)dv$. For large $m$, the time
scale $\delta_m=m^{-2}$ and the functions
$f_m(u)=\min\{u/\delta_m,1\}$ and
$h_m(u)=[mf_m(u)-b(u)]/\sigma$ define the event
\[
 E_m=\left\{\sup_{0\leq u\leq T}|W_u-h_m(u)|\leq\epsilon/\sigma\right\}.
\]
On $E_m$,
\[
 \int_0^T|\bar S_T-\bar S_u|du
 \leq 2se^m(e^\epsilon\delta_m+T\sinh\epsilon).
\]
For the geometric option, $r\geq0$ and
$\int_0^T f_m(u)du=T-\delta_m/2$ imply
\[
\frac{R_T}{T}\geq\log s+m-\frac{1}{2Tm}-\epsilon.
\]
Hence, for sufficiently large $m$, its geometric average is at least
$se^{m-\epsilon-1}$. The arithmetic average is no smaller, so for
either payoff,
\[
 n\Phi(R_T)\geq ns e^{m-\epsilon-1}-nK.
\]
For sufficiently small $\epsilon>0$ and large $m$, it follows from
\eqref{eq:zero-inventory-wealth-bound} that
$W_T^{+1,\nu}\leq C-ce^m$ on $E_m$, with $c>0$ independent of the
strategy.

The shift $h_m$ is absolutely continuous, with $h_m(0)=0$ and
$h_m'\in L^2([0,T])$. The event
$\{\sup_{0\leq u\leq T}|W_u|\leq\epsilon/\sigma\}$ has probability
$p_\epsilon>0$ and is invariant under $W_u\mapsto-W_u$.
From the Cameron--Martin formula
\citep[Theorem 1.38 and Remark 1.43]{MortersPeres2010}, symmetry
and conditional Jensen's inequality,
\[
\mathbb P(E_m)
 \geq p_\epsilon\exp\left\{-\frac12\int_0^T|h_m'(u)|^2du\right\}
 \geq c_1e^{-c_2(1+m^4)},
\]
since $\int_0^T|h_m'(u)|^2du=O(m^4)$.
Consequently,
\[
 \mathbb E^{\mathbb P}[e^{-\gamma W_T^{+1,\nu}}]
 \geq c_1\exp\{\gamma ce^m-\gamma C-c_2(1+m^4)\}
 \longrightarrow\infty.
\]
The bound is uniform over admissible strategies and probability
spaces. With no option, the zero trading rate implies
\[
u^0(0,x,s,i,0,0,0)\geq-e^{-\gamma e^{rT}x}>-\infty.
\]
Every finite initial payment leaves the seller's utility value equal
to $-\infty$, so no finite payment satisfies
\eqref{eq:dealer-ask-definition}.
\end{proof}

\end{document}